%% file: letter.tex
\documentclass[aps,prl, reprint,
amsmath,amssymb,nofootinbib,longbibliography]{revtex4-1}
\usepackage{graphicx}
\usepackage[dvipsnames,table]{xcolor}
\usepackage{hyperref}
\usepackage{xspace}
\usepackage{orcidlink}
\usepackage[caption=false]{subfig}
\usepackage{slashed}
\usepackage{cleveref}

\colorlet{blobcolor}{CadetBlue}

\colorlet{n1-arrow}{ProcessBlue}
\colorlet{n2-arrow}{Plum}
\colorlet{n3-arrow}{orange}

\usepackage{tikz}
\input{tikz_preamble}

\begin{document}
\author{John Joseph M. Carrasco  \orcidlink{0000-0002-4499-8488}}
\affiliation{Department of Physics and Astronomy, Northwestern
  University, Evanston, Illinois 60208, USA}
\author{Alex Edison \orcidlink{0000-0002-5430-9500}}
\affiliation{Department of Physics and Astronomy, Northwestern
  University, Evanston, Illinois 60208, USA}
\author{Nia Robles Del Pino \orcidlink{0009-0008-6708-0099}}
\affiliation{Department of Physics and Astronomy, Northwestern
  University, Evanston, Illinois 60208, USA}
\author{Suna Zekio\u{g}lu \orcidlink{0000-0002-5019-7310}}
\affiliation{Department of Physics and Astronomy, Northwestern
  University, Evanston, Illinois 60208, USA}

\title{An exercise in Color-Dual Cut Tiling: \\
$\mathcal{N}=8$ Supergravity from Positivity}

\begin{abstract}
  The BCJ duality between color and kinematics brings two advantages
  to calculating multi-loop scattering amplitudes.  First the number
  of ordered cuts that need to be performed to fix the integrand to a
  gauge theory is minimal -- reducing the factorially-growing number
  of diagrams to a single-digit basis in all known cases. Second is
  the trivial generation of related gravitational amplitudes from
  gauge theory amplitudes via double-copy.  Mounting evidence suggests
  there are some cases where no local color-dual representations
  exist, even when the semi-classical theory is color-dual.  Can we
  still simplify the calculations without making the duality manifest
  at the level of the integrand?  Here we take a non-trivial step in
  this direction by showing that the satisfaction of tree-level BCJ
  relations is sufficient to dramatically reduce the number of
  explicit cuts that must be performed, even when the loop-level
  relations are not explicitly satisfied.  We introduce an
  agglomerative algorithm, color-dual cut tiling, that identifies and
  builds the entire integrand from the simplest on-shell conditions
  applied to a seed of off-shell integrand information.  Specifically,
  we demonstrate that for two-to-two scattering at three loops in the
  maximally supersymmetric gauge theory there is sufficient
  information contained in planar cuts — completely determined by
  positivity constraints — to generate all of the non-planar sector.
  Additionally, we make use of the generalized double copy to generate
  a representation of maximally supersymmetric gravity as a functional
  of the planar SYM input.  We discuss how the process might
  generalize, and then close by commenting on the applicability of
  this method for additional cases of interest where performing
  explicit unitarity cuts is expensive.
\end{abstract}

\maketitle

\section{Introduction}

The computational complexity of extracting physical prediction from
relativistic quantum field theory poses a fundamental challenge.  When
the coupling is relatively small for the scales at hand, then
perturbative expansion around the interaction strength can yield a
controlled approximation.  Even here the factorial growth in
complexity order by order in interaction strength places hard
constraints on what can be knowable with finite resources — yet the
universe appears tractable. Gauge-invariant physical on-shell
quantities collapse to rather compact expressions, despite the
exhaustively verbose intermediary steps encountered using traditional
off-shell Feynman methods.  This realization has energized the field
of scattering amplitudes to develop a host of on-shell methods all
stemming from applications of unitarity cuts \cite{Bern:1994zx,
  Bern:1994cg, Britto:2004nc, Anastasiou:2006jv, Bern:2007ct,
  Carrasco:2023qgz, Travaglini:2022uwo, Bern:2022wqg,
  Carrasco:2021otn, Bern:2018jmv, Bern:2012uf, Edison:2022smn,
  Bourjaily:2020qca, Bourjaily:2017wjl, Arkani-Hamed:2012zlh}. In a
myriad of theories, from the formal to the phenomenologically
effective, on-shell methods have time and again proven extremely
efficient at extracting predictions from local quantum field theory,
even providing suggestions of the quantum nouns and verbs necessary to
ultimately go beyond it.

Idealized systems, such as the four-dimensional $\mathcal{N}=4$
super-Yang Mills (SYM) theory and the closely related maximal
$\mathcal{N}=8$ supergravity (SG) theory, have been mined for
structure.  The symmetries of these idealized theories admit a
technical simplicity for calculation that lets us probe higher orders
in coupling, or loop level, than in their less symmetric, more
phenomenological cousins.  Famously, construction of the Amplituhedron
demonstrates that the planar limit of $\mathcal{N} = 4$ SYM, including
multi-loop integrands, is the unique answer to concrete questions of
positivity \cite{Arkani-Hamed:2013jha, Arkani-Hamed:2014dca}.
Building on the discovery of dual conformal invariance (DCI) in
maximal SYM \cite{Drummond:2008vq}, the study of planar amplitudes has
found great success in pushing both conceptual and practical
understanding of scattering \cite{Bourjaily:2016evz,
  Arkani-Hamed:2012zlh, Arkani-Hamed:2013jha, Arkani-Hamed:2014dca,
  Bern:2012di, Dixon:2021nzr, Caron-Huot:2019vjl, Drummond:2008vq}.

For gauge theories, the fact that adjoint color weights of distinct
graphs can be related through Jacobi identities allows for
cancellations of gauge dependence between Feynman graphs.  First
observed in $\mathcal{N}=4$ SYM~\cite{Bern:2008qj}, the duality
between color and kinematics asserts the existence of a representation
where kinematic numerators of Feynman graphs obey they same algebraic
relations as the color weights.  Double-copy construction maps vector
amplitudes to gravity amplitudes by replacing color weights with
color-dual gauge numerators, promoting linearized gauge invariance to
linearized diffeomorphism invariance. By relating gauge and gravity
predictions directly graph by graph, this clarified aspects of the
tree-level relation between gravitational scattering amplitudes and
color-stripped gauge theory amplitudes discovered as a low-energy
limit of string theory amplitudes \cite{Kawai:1985xq}. Both the
duality between color and kinematics and associated double-copy
construction were rapidly generalized to integrands at
multi-loop-level \cite{Bern:2010ue, Bern:2017yxu}, a wide web of
color-dual double-copy theories \cite{Cachazo:2014xea,Bern:2019prr,
  Adamo:2022dcm}, and classical solutions \cite{Bern:2019prr,
  Adamo:2022dcm,Monteiro:2014cda, Kosower:2018adc, Buonanno:2022pgc}.

Unfortunately loop-level color-dual constraints must be functional,
which necessitates increasingly large ansatze at higher loop level.
There have even been suggestions that, in certain cases, local
color-dual representations may not even exist \cite{Bern:2015ooa,
  Mogull:2015adi, Edison:2023ulf, Li:2023akg}. However, even in the
absence of loop-level color-dual gauge theory representations, we may
still generate gravity amplitudes via generalized double
copy~\cite{Bern:2017yxu} by double-copying violations of kinematic
Jacobi identities to dress contact graphs. Hence, it is still quite
advantageous to be able to write down loop-level gauge theory
integrands even if a local color-dual representation does not appear
possible.

For all the advantages that unitarity methods offer over traditional
integrand construction via Feynman rules, cut construction is not
free. At higher orders, or when large gamma-traces come into play, the
computational complexity of calculating individual cuts can be quite
prohibitive \cite{Carrasco:2021otn}.  Here lies one of the advantages
of finding color-dual numerators at the multi-loop level:
systematically relating the numerators of graphs ultimately minimizes
the number of unitarity cuts that need to be performed.

In this letter, we demonstrate that the existence of
\textit{tree-level} color-kinematic relations still allows us to
dramatically reduce the number of ordered cuts that must be performed
in order to generate the full gauge theory integrand.  We introduce an
agglomerative method we call color-dual cut tiling.  It starts with
seeded knowledge of certain topologies then systematically identifies
the simplest on-shell conditions that can be applied to those
topologies to push that data to the rest of the integrand without
evaluating cuts by sewing trees together.  The seeded knowledge could
come from explicit cut evaluations to tree-level or via principles
that hold in a certain well-defined sector of the integrand.  We push
out that seeded data by exploiting a corollary of the color-kinematics
duality: the tree-level $(n-3)!$ BCJ amplitude relations
\cite{Bern:2008qj}.  These relations can be derived in a number of
ways \cite{Bern:2008qj, Cachazo:2013hca, Carrasco:2016ldy}, but can be
summarized as a minimal basis statement for colored-ordered
amplitudes:
\begin{equation}
  A_{\text{tree}}(\sigma(1) \dots \sigma(n))
  = \sum_{\tau \in S_{n-3}} C_{\sigma | \tau} A_{\text{tree}}(1,2,3,\tau)
  \label{eq:nm3-rel}
\end{equation}
where the $C_{\sigma | \tau}$ are rational functions of the kinematic
Mandelstam variables.  The idea is that if certain orderings can be
dressed in terms of exact knowledge of integrands under appropriate on
shell conditions, then these relations can be used to dress other
sectors of the integrand by simply multiplying and dividing by momentum
invariants as we describe.  We demonstrate the power of this method by
answering the sharp question: can knowledge of planar maximally
supersymmetric Yang-Mills at a given loop order and multiplicity be
sufficient to construct (non-planar) gravity amplitudes in
four-dimensions?
 
As described earlier, positivity constrains the planar sector of
$\mathcal{N}=4$ SYM to be dual conformally invariant.  However, beyond
two loops, dual-conformal numerators are inconsistent with manifesting
the duality between color and kinematics, preventing the
\text{na\"ive} tiling of simply applying Jacobi moves off shell.  In
the case of maximally supersymmetric Yang-Mills in the planar limit,
one can construct the planar integrand via positivity
\cite{Arkani-Hamed:2013jha, Arkani-Hamed:2014dca}.  In this letter we
will show that we can use color-dual cut tiling to push such
information out at three-loops, fully reconstructing the
${\mathcal N}=8$ SG integrand at three-loops without evaluating a
single cut in terms of sewing tree amplitudes -- only by manipulation
of integrands under on-shell conditions.  This interesting result
opens the question: for what amplitudes in gauge theories is the
planar sector sufficient to completely construct the non-planar
integrand?  We present a general analysis for four points through four
loops and discuss future calculations of interest where these ideas
could prove useful.

\section{Color-Dual Cut Tiling}

We begin by introducing a few definitions.  A \textit{cut topology}
can be described by a multi-node graph, with nodes representing
tree amplitudes and edges representing on-shell conditions.  Any
\textit{color-ordered cut} can be represented by a particular cut
topology with additional specification of order around each
vertex. For gauge theories, a cut topology can be thought of as
representing a family of color-ordered cuts.

Color-ordered cuts can be mapped to a sewing operation on trees where
each vertex is understood to represent a tree-level amplitude: one
takes the product of all trees and sums over the physical states of
the theory for all shared internal edges.  We refer to that operation
as \textit{evaluating a cut.}  Depending on the theory, there may be a
number of physical states that contribute to each cut.

We can also impose cut conditions on our candidate loop-level
integrands for the theory. This is done by identifying all Feynman
graphs that contribute to the cut (those that can be labeled in such a
way as to share on-shell conditions with a particular ordered cut),
and then writing down the contribution from each graph using the
candidate loop-level integrand representation. We refer to this
procedure of applying an integrand to a cut as \textit{dressing a
  cut.} If only planar graphs contribute to a particular ordering of a
cut topology, we call that ordered cut planar.

Since we may evaluate cuts with our tree-level (color-dual) knowledge
of the theory, we can match with dressed cuts to fix our loop-level
integrands. The only contributions that might be missing are contact
terms that vanished under the cut conditions.  A \textit{spanning set
  of cuts} for a theory is a collection of cuts that ensures there are
no missing contacts.

The $(n-3)!$ basis of \cref{eq:nm3-rel} allows us to relate different
orderings of tree-level amplitudes; we can therefore relate different
cut orderings to one another within the same cut topology. If all
associated ordered cuts may be expressed solely in terms of a basis of
planar cut orderings, we call that cut topology \textit{planar
  dominated}, as it is spanned by planar data. It is necessary that a
cut topology's graph be planar for it to be planar dominated, but this
is not a sufficient criteria.

How are minimal spanning cuts identified?  In the method of maximal
cuts (MMC)~\cite{Bern:2007ct} one builds up kinematic weights
associated with cubic graphs contact term by contact term.  As contact
terms vanish under cut conditions, one first finds the indelible
information associated with a graph by cutting all of its
propagators. For cubic (or trivalent) graph representations, this
means only sewing together three-point trees. Such cuts of cubic
graphs are called {\it maximal cuts}.  Releasing cut conditions
(merging adjacent trees to higher-point trees) exposes any remaining
contact terms. These can be assigned either to higher valence
topologies or associated with the cubic graphs that can contribute to
that less than maximal cut. We refer to cuts with one (many) released
cut conditions as next-to-maximal (N${}^k$Max) cuts.  The
power-counting of a theory determines the number of cut conditions
that must ever be released -- although of course nothing prevents us
from releasing additional cut constraints towards verification of a
newfound integrand.

Our color-dual cut tiling approach applies a similar logic, but with
the critical restriction to only considering planar-dominated cut
topologies.  We begin by assigning the known planar representation to
our planar graphs and vanishing kinematic weight to all non-planar
graphs. Starting at the next-to-maximal cut level, we identify which
planar-dominated cut topologies are not satisfied given the current
dressings -- those are the cut topologies which provide novel insight
into the non-planar ramifications of the planar sector.  We amend our
non-planar kinematic dressings accordingly to satisfy these cuts, and
then consider the next level of planar dominated cuts to identify any
missing information.  This approach is agglomerative in the sense that
it considers only the simplest unresolved planar-dominated cuts
necessary to provide data that have not been assigned by simpler cuts,
order by order in released cut conditions.  It merges the process of
identifying relevant planar-dominated cuts with the process of
cut-information assignment, starting with the
simplest\footnote{Minimizing the number of contributing graphs to each
  cut.} cuts.  This need not be the case as we discuss shortly.

While it is certainly possible to give each non-planar graph a minimal
power-counting ansatz and fix to these cut conditions, in this case it
is not necessary. The cuts identified by our tiling scheme can be
directly used to fix the non-planar dressings. Each non-planar cut
ordering will have kinematic data sitting on off-shell propagators
that we can thereby associate with particular channels. Such data is
directly assigned to the relevant non-planar graph's kinematic
contribution. Any kinematic numerator associated with a graph is
required to obey the graph's automorphisms, and so any such assignment
is symmetrized.  There is freedom when assigning contact terms to
cubic graphs, but any realization of that freedom that satisfies
automorphism invariance is sufficient.

\begin{figure}
  \begin{tikzpicture}
          \coordinate (shiftr) at (2,0);
      \coordinate (shiftv) at (0,1);
    \begin{scope}
      \pic{diag a={0.8}};
    \end{scope}
    \begin{scope}[shift={($1.2*(shiftr)-(shiftv)$)}]
      \pic{box skel gen={0.8}{1}{0}{2}{0}};
      \draw (vru) -- ++(45:\extL/2);
      \draw (vrd) -- ++(-45:\extL/2);
      \draw (vlu) -- ++(135:\extL/2);
      \draw (vld) -- ++(225:\extL/2);
      \draw (mu1) -- (md2);
      \draw (vru) -- (md1);
      
      \draw (vlu) -- (vru) -- (vrd) -- (vld) -- (vlu);
      \node[small blob] at (vru) {};
      \coordinate(c2r) at (0:0.8);
      \coordinate(c2l) at (180:0.8);
      \coordinate(c2d) at (270:0.8);
      \node[] at (90:0.8){\(a_{1;2} \)};
      \coordinate (dbound) at ($(vld)+(225:\extL/2)$);
    \end{scope}
    
    \begin{scope}[shift={($1.2*(shiftr)+(shiftv)$)}]
      \pic{box skel gen={0.8}{2}{0}{1}{0}};
      \draw (vru) -- ++(45:\extL/2);
      \draw (vrd) -- ++(-45:\extL/2);
      \draw (vlu) -- ++(135:\extL/2);
      \draw (vld) -- ++(225:\extL/2);
      \draw (mu1) -- (md1);
      \draw (mu2) -- (md1);
      \draw (vlu) -- (vru) -- (vrd) -- (vld) -- (vlu);
      \node[small blob] at (md1) {};
      \coordinate(c1r) at (0:0.8);
      \coordinate(c1l) at (180:0.8);
      \coordinate (ubound) at ($(vlu)+(135:\extL/2)$);
      \node[] at (90:0.8){\(a_{1;1} \)};
    \end{scope}

    \begin{scope}[shift={($2.2*(shiftr)$)}]
      \pic{box skel gen={0.8}{2}{0}{0}{0}};
      \draw (vru) -- ++(45:\extL/2);
      \draw (vrd) -- ++(-45:\extL/2);
      \draw (vlu) -- ++(135:\extL/2);
      \draw (vld) -- ++(225:\extL/2);
      \draw (mu1) -- (vld);
      \draw (mu2) -- (vrd);
      
      \draw (vlu) -- (vru) -- (vrd) -- (vld) -- (vlu);
      \node[small blob] at (vrd) {};
      \node[small blob] at (vld) {};
      \coordinate(c3d) at (270:0.8);
      \node[] at (90:0.8){\(a_{2;1} \)};
    \end{scope}
    \begin{scope}[decoration={calligraphic brace,amplitude=6pt}]
      \draw[ultra thick,decorate]($(-0.5,0)+(dbound)$)--($(-0.5,0)+(ubound)$);
    \end{scope}
    \begin{scope}[shift={($-3.5*(shiftv)+1.2*(shiftr)$)}]
    \coordinate(bu) at (90:0.8);
    \pic{diag b={0.8}};
    \end{scope}y
    \begin{scope}[shift={($-3.5*(shiftv)+0.2*(shiftr)$)}]
      \coordinate(cu) at (90:0.8);
      \pic{diag c={0.8}};
    \end{scope}
    \begin{scope}[shift={($-3.5*(shiftv)+2.2*(shiftr)$)}]
      \coordinate(du) at (90:0.8);
      \pic{diag d={0.8}};
    \end{scope}

    \draw[thick,n2-arrow,-Stealth] (c3d) to (du);
    \draw[thick,n2-arrow,-Stealth] (c3d)
    .. controls ($(c3d)-2.5*(shiftv)$) and ($(bu) + (shiftv)+(.4,0)$) ..
    ($(bu)+(0.2,0)$);
    \draw[thick,n1-arrow,-Stealth] (c2d) to[out=270,in=90] (bu);
    \draw[thick,n1-arrow,-Stealth] (c1l) to [out=220,in=90] ($(c1l)+(260:2)$)
    .. controls ($(c1l)+(260:2)+(0,-2.3)$) and ($(cu)+(0,1)$) .. (cu);
 \end{tikzpicture}
  
 \caption{The three planar dominated cut-topologies $a_{1;1}$,
   $a_{1;2}$, and $a_{2;1}$ --- whose ordered cuts are spanned by
   knowledge of the planar (a) graph contribution --- are entirely
   sufficient to specify the kinematic numerators of graphs (c), (b),
   and (d) for the $\mathcal{N}$=4 SYM theory. Each arrow from a cut
   topology $a_{k;i}$ to a non-planar graph $(x)$ represents an
   equation from an ordering that relates cut data supplied by graph
   (a) with the on-shell conditions of the cut and at least one
   labeling of the non-planar graph $(x)$.  Merged arrows mean an
   equation with more than one non-planar graph contribution. The
   ordering of every four-point vertex determines which two of three
   cubic sub-graphs contributes, and the ordering around every
   five-point vertex determines which five of fifteen possible cubic
   sub-graphs contributes.  The no-triangle property of
   $\mathcal{N}=4$ SYM excludes any need for non-vanishing triangle or
   bubble sub-graphs.  }
\label{aDominated}
\end{figure}
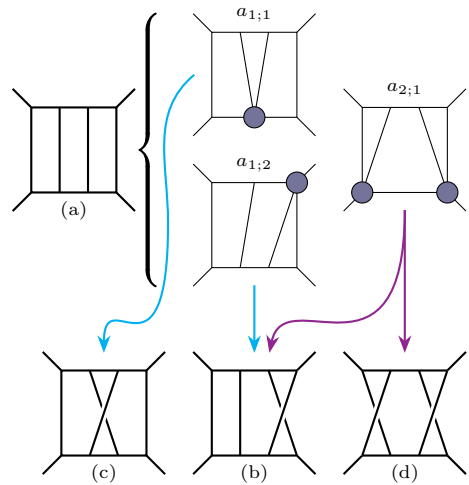

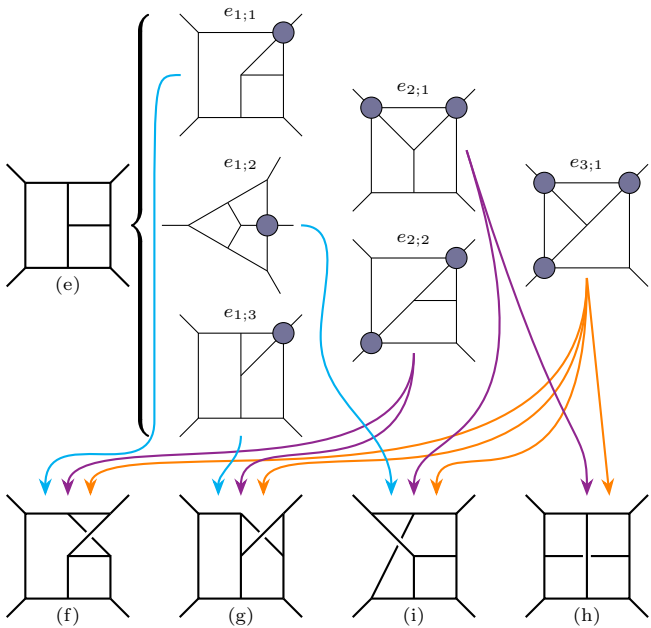
\begin{figure}
  \begin{tikzpicture}
    \coordinate (shift1) at (2.3,0);
    \coordinate (shift2) at (0,-2);
    \coordinate (initshift) at (0.3,0);
    \begin{scope}
      \pic{diag e={0.8}};
    \end{scope}
    \begin{scope}[decoration={calligraphic brace,amplitude=6pt}]
      \draw[ultra thick,decorate]
      ($(mr1)+(0.5,0)+1.4*(shift2)$)--($(mr1)+(0.5,0)-1.4*(shift2)$);
    \end{scope}
    
    \begin{scope}[shift={($(shift1)+(shift2)$)}]
      \pic{box skel gen={0.8}{1}{0}{1}{0}};

      \draw (md1)-- (mid) -- (vru);
      \draw (mid) -- (mu1);
      
      \draw (vru) -- ++(45:\extL/2);
      \draw (vrd) -- ++(-45:\extL/2);
      \draw (vlu) -- ++(3*45:\extL/2);
      \draw (vld) -- ++(-3*45:\extL/2);
      \draw (vlu) -- (vru) -- (vrd) -- (vld) -- (vlu);
      \node[small blob] at (vru){};
      \coordinate(c3r) at (0:0.8);
      \coordinate(c3l) at (180:0.8);
      \coordinate(c3d) at (270:0.8);
      \node[] at (90:0.8){\(e_{1;3} \)};
    \end{scope}

    \begin{scope}[shift={($(shift1)-(shift2)$)}]
      \pic{box skel gen={0.8}{0}{1}{1}{0}};

      \draw (md1)-- (mid) -- (vru);
      \draw (mid) -- (mr1);
      
      \draw (vru) -- ++(45:\extL/2);
      \draw (vrd) -- ++(-45:\extL/2);
      \draw (vlu) -- ++(3*45:\extL/2);
      \draw (vld) -- ++(-3*45:\extL/2);
      \draw (vlu) -- (vru) -- (vrd) -- (vld) -- (vlu);
      \node[small blob] at (vru){};
      \coordinate(c1r) at (0:0.8);
      \coordinate(c1l) at (180:0.8);
      \node[] at (90:0.8){\(e_{1;1} \)};
    \end{scope}

    \begin{scope}[shift={($(shift1)$)}]
      \pic{tri skel={0.7}};   
      \foreach
      \tv [remember=\tv as \lasttv (initially ml)] in {vl,mu,vu,mr,vd,ml}{
       
        \draw (\lasttv) -- (\tv);
      }
      \foreach \tv in {mu,mr,ml}{
        
        \draw (mid) -- (\tv);
      }   
      
      \draw (vu) -- ++(60:\extL/2);
      \draw (vd) -- ++(-60:\extL/2);
      \draw (vl) -- ++(180:\extL/2);
      \draw(mr) -- ++(0:\extL/2);
      \node[small blob] at (mr){};
      \coordinate(c2r) at (0:0.8);
      \coordinate(c2l) at (180:0.7);
      \node[] at (90:0.8){\(e_{1;2} \)};
    \end{scope}
  
    \begin{scope}[shift={($2*(shift1)-0.5*(shift2)$)}]
           \pic{box skel={0.8}};
     \draw (vlu) -- (vru) -- (vrd) -- (vld) -- (vlu);
     \draw (vlu) -- (mid) -- (vru);
     \draw (mid) -- (md);

     \draw (vru) -- ++(45:\extL/2);
     \draw (vrd) -- ++(-45:\extL/2);
     \draw (vlu) -- ++(3*45:\extL/2);
     \draw (vld) -- ++(-3*45:\extL/2);
     \node[small blob] at (vru){};
     \node[small blob] at (vlu){};
     \coordinate(c5r) at (0:0.7);
     \coordinate(c5l) at (180:0.7);
     \coordinate(c5d) at (270:0.7);
     \node[] at (90:0.8){\(e_{2;1} \)};
   \end{scope}
   \begin{scope}[shift={($2*(shift1)+0.5*(shift2)$)}]
     \pic{box skel gen={0.8}{0}{1}{1}{0}};
     \draw (vlu) -- (vru) -- (vrd) -- (vld) -- (vlu);
     \draw (vld) -- (mid) -- (vru);
     \draw (mid) -- (mr1);

     \draw (vru) -- ++(45:\extL/2);
     \draw (vrd) -- ++(-45:\extL/2);
     \draw (vlu) -- ++(3*45:\extL/2);
     \draw (vld) -- ++(-3*45:\extL/2);
     \node[small blob] at (vru){};
     \node[small blob] at (vld){};
     \coordinate(c4r) at (0:0.7);
     \coordinate(c4l) at (180:0.7);
     \coordinate(c4d) at (270:0.7);
     \node[] at (90:0.8){\(e_{2;2} \)};
   \end{scope}

   \begin{scope}[shift={($3*(shift1)$)}]
          \pic{box skel={0.8}};
     \draw (vlu) -- (vru) -- (vrd) -- (vld) -- (vlu);
     
     \draw (mid) -- (vlu);
     \draw (mid) -- (vru);
     \draw (mid) -- (vld);
     \draw (vru) -- ++(45:\extL/2);
     \draw (vrd) -- ++(-45:\extL/2);
     \draw (vlu) -- ++(3*45:\extL/2);
     \draw (vld) -- ++(-3*45:\extL/2);
     \node[small blob] at (vru){};
     \node[small blob] at (vlu){};
     \node[small blob] at (vld){};
     \coordinate(c6d) at (270:0.7);
     \node[] at (90:0.8){\(e_{3;1} \)};
   \end{scope}
   
   \begin{scope}[shift={($2.2*(shift2)$)}]
     \coordinate(fu) at (90:0.8);
     \pic{diag f={0.8}};
   \end{scope}
   \begin{scope}[shift={($(shift1)+2.2*(shift2)$)}]
     \coordinate(gu) at (90:0.8);
     \pic{diag g={0.8}};
   \end{scope}
   \begin{scope}[shift={($3*(shift1)+2.2*(shift2)$)}]
     \coordinate(hu) at (90:0.8);
     \pic{diag h={0.8}};
   \end{scope}
   \begin{scope}[shift={($2*(shift1)+2.2*(shift2)$)}]
     \coordinate(iu) at (90:0.8);
     \pic{diag i={0.8}};
   \end{scope}

   \foreach \td in {fu,gu,iu}{
     \draw[thick,n3-arrow,-Stealth] (c6d) 
     .. controls ($(c6d)+(270:3.5)$) and ($(\td)+(0.3,1)$) ..
     ($(\td)+(0.3,0)$);
  }
  \draw[thick,n3-arrow,-Stealth] (c6d) -- ($(hu)+(0.3,0)$);
  
  \foreach \td in {fu,gu}{
    \draw[thick,n2-arrow,-Stealth] (c4d) 
    .. controls ($(c4d)+(270:2)$) and ($(\td)+(0,1)$) ..
    ($(\td)+(0,0)$);
  }
    \foreach \td in {hu,iu}{
    \draw[thick,n2-arrow,-Stealth] (c5r) 
    .. controls ($(c5d)+(-60:4)$) and ($(\td)+(0,1)$) ..
    ($(\td)+(0,0)$);
  }

  \draw[thick,n1-arrow,-Stealth] (c3d) to[out=270,in=90] ($(gu)-(0.3,0)$);
  
  \draw[thick,n1-arrow,-Stealth] (c2r) to[out=0,in=90] ($(c2r)+(280:1.5)$)
  to [out=270,in=90]($(iu)-(0.3,0)$);

  \draw[thick,n1-arrow,-Stealth] (c1l)
  ..controls($(c2l)+(180:0.45)+(0,2)$) ..
  ($(c2l)+(180:0.45)$) -- ($(c2l)+(180:0.45)+(270:2.6)$)
  .. controls ($(c2l)+(180:0.45)+(270:3.5)$) and ($(fu)-(0.3,-1)$) ..
  ($(fu)-(0.3,0)$);
    
  \end{tikzpicture}
  
  \caption{Knowledge of only the five planar dominated cut-topologies
    labeled $e_{k;i}$ --- whose ordered cuts are spanned by knowledge of
    the planar (e) graph contribution --- is sufficient to construct
    the kinematic numerators of graphs (f), (g), (h), and (i) for the
    $\mathcal{N}$=4 SYM theory.  Each arrow from a cut topology
    $e_{k;i}$ to a non-planar graph $(x)$ represents an equation from
    an ordering that relates cut data supplied by graph (e) with the
    on-shell conditions of the cut and at least one labeling of the
    non-planar graph $(x)$.  Merged arrows mean an equation with more
    than one non-planar graph contribution. }
\label{eDominated}
\end{figure}

At three loops, in the maximally supersymmetric gauge theory, eight
simple planar-dominated cut topologies are sufficient to generate the
entire integrand from the two planar cubic graphs (labeled $(a)$ and
$(e)$). In \cref{aDominated} and \cref{eDominated}, we illustrate how
these cut topologies map data from the planar graphs to the non-planar
sector.

For the NMax cut topology $a_{1,1}$, one may identify both a cut
ordering to which only $(a)$ contributes and a cut ordering to which
only $(c)$ and $(a)$ contribute. Since this cut topology only involves
one four-point tree, we need only one ordering to be planar for it to
be planar dominated. By relating these orderings using the $(n-3)!$
relations, we begin building up our dressing of the $(c)$ graph. This
turns out to be entirely sufficient to identify the full contribution
of the non-planar graph $(c)$. The combination of cut topologies
$a_{1;2}$ (NMax) and $a_{2;1}$ (N${}^2$Max) are sufficient to entirely
inform the contributions of graphs $(b)$ and $(d)$.  Non-planar graphs
$(f)$-$(h)$ are entirely constrained by the five planar-dominated cut
topologies from NMax ($e_{1;1}, e_{1;2}, e_{1;3}$), N${}^2$Max
($e_{2;1}, e_{2;2}$), and N${}^3$Max ($e_{3;1}$) levels, as
demonstrated by \cref{eDominated}.

Since, as we will show below, the spanning four-particle cut topology
between two six-point trees is planar-dominated, we can verify that we
have constructed the entire integrand, without actually sewing any
trees together, by recycling exact knowledge of the planar sector.
The resulting gravitational amplitude generated using generalized
double copy of course satisfies all $\mathcal{N}=8$ SG cuts.

\section{Constructing non-planar integrands from only planar data}

The prescriptive cut-tiling approach described above is an
agglomerative approach that mixes the identification of a spanning set
of minimally complicated planar-dominated cuts with the assignment of
cut data to non-planar graphs.  This need not be the case.
Identification of a spanning set of planar-dominated cuts can be
disentangled from assignment as we describe here.  Indeed, the task of
identifying whether or not there is sufficient planar data in any
gauge theory can be reduced to a simple question: is the minimal set
of spanning cuts of an integrand planar dominated?  If so, then planar
data is all that is required to construct the entire non-planar
amplitude.  With a spanning set of unitarity cuts in hand, it is
possible to reconstruct a complete integrand using any cut merging
algorithm \cite{Bern:2007ct, Bern:2012uf, Bern:2015ooa,
  Bourjaily:2017wjl}.  Furthermore, one can generate gravitational
amplitudes from a complete set of gauge theory numerators via the
generalized double copy.  As such, a planar-spanned basis of cuts
implies that the planar sector of the theory encodes the entirety of
both the non-planar sector as well as its contributions to a
corresponding gravitational double copy.

We will now demonstrate that both 2- and 3-loop 4-point amplitudes for
any gauge theory have a planar-spanned basis of cuts.  (The 4-point
1-loop amplitude is trivially planar-dominated, as every diagram is
planar with respect to some external ordering.)  We will also offer
comments on the boundaries of the process with respect to higher
loops.

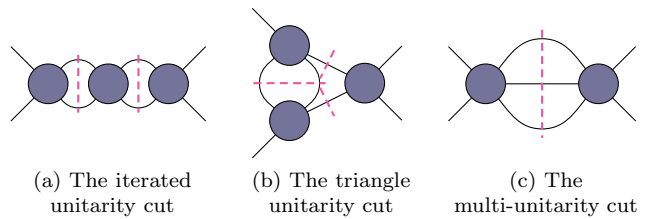
\begin{figure}[ht]
  \subfloat[The iterated unitarity cut]{
    \begin{tikzpicture}
      \coordinate (a1) at (0,0);
      \coordinate (a2) at (0.8,0);
      \coordinate (a3) at (1.6,0);
      \draw pic {muc int={2}{a1}{a2}};
      \draw pic {muc int={2}{a2}{a3}};
      \draw pic {exts={a1}{a3}};
      \node[blob] at (a1) {};
      \node[blob] at (a2) {};
      \node[blob] at (a3) {};
      \draw pic {muc strut};
     \coordinate (s1) at (0,-0.5);
     \coordinate (s2) at (0,0.5);
     \path (s2) -- ++(135:0.7);
      \path (s1) -- ++(225:0.7);
    \end{tikzpicture}
    \label{fig:two-unit}
  } \hfill
  \subfloat[The triangle unitarity cut]{
    \begin{tikzpicture}
      \coordinate (a1) at (0,0);
      \coordinate (a2) at (0,1);
      \coordinate (a3) at (1,0.5);
      \draw pic {muc int={2}{a1}{a2}};
      \draw pic {extr={a3}};
      \draw (a2) -- ++(135:0.7);
      \draw (a1) -- ++(225:0.7);
      \draw pic {cut line={a2}{a3}{0.3}};
      \draw pic {cut line={a1}{a3}{0.3}};
      \node[blob] at (a1) {};
      \node[blob] at (a2) {};
      \node[blob] at (a3) {};
    \end{tikzpicture}
    \label{fig:two-tri}
  } \hfill
   \subfloat[The multi-unitarity cut]{
     \begin{tikzpicture}
       \draw pic {muc={3}};
       \coordinate (s1) at (0,-0.5);
       \coordinate (s2) at (0,0.5);
       \path (s2) -- ++(135:0.7);
       \path (s1) -- ++(225:0.7);
    \end{tikzpicture}
    \label{fig:two-muc}
   }
  \caption{The spanning physical generalized unitarity cuts for 2-loop
    4-point.}
\label{fig:two-spanning}
\end{figure}

For the two-loop four-point amplitude, we need to evaluate the set of
physical spanning generalized unitarity cuts shown in
\cref{fig:two-spanning}.  All three of these cut topologies can be
fully reconstructed, including their non-planar contributions, in
terms of purely planar input.

We start with the iterated unitarity cut topology,
\cref{fig:two-unit}.  This cut (and any continued iteration) is
actually fully expressed in terms of planar diagrams: Inserting a
$u$-channel expansion for any of the blobs is equivalent to a
relabeling of a planar diagram.

Next the ``generalized triangle'' cut topology, \cref{fig:two-tri}.
Associated ordered cuts contain non-planar channels; for instance, the
``cross-box'' cubic diagram results from inserting a $u$-channel cubic
expansion into each of the left two blobs, and a $t$-channel expansion
into the right.  However, because all of the constituent amplitudes
are 4-point, \cref{eq:nm3-rel} applied to each amplitude allows us to
always express all of the color channels in terms of the $st$-channel
expansion which produces planar diagrams.

\begin{figure}[ht]
\begin{equation*}
  \left(
    \begin{array}{c}
      \begin{tikzpicture}
        \pgfmathsetmacro{\fivesep}{0.4}

        \coordinate(i1) at (0,\fivesep);
        \coordinate(i2) at (0,0);
        \coordinate(i3) at (0,-\fivesep);
        \coordinate(cp) at ($(i1)!0.5!(i3)$);
        
        \node[left] (e1) at ($(i1) - (\fivesep,0)$){2};
        \node[left](e2) at ($(i3) - (\fivesep,0)$){1};
        
        \node[right] (e3) at ($(i1) + (\fivesep,0)$) {\(\ell_1\)};
        \node[right] (e4) at ($(cp) + (\fivesep,0)$){\(\ell_2\)};
        \node[right](e5) at ($(i3) + (\fivesep,0)$){\(\ell_3\)};
        \draw (e1) -- (i1) -- (e3);
        \draw (e2) -- (i3) -- (e5);
        \draw (i2) -- (e4);
        \draw (i1) -- (i3);
      \end{tikzpicture}
      \\
      \begin{tikzpicture}
        \pgfmathsetmacro{\fivesep}{0.4}
        
        \coordinate(i1) at (0,\fivesep);
        \coordinate(i2) at (0.3*\fivesep,\fivesep);
        \coordinate(i3) at (0,-\fivesep);
        \coordinate(cp) at ($(i1)!0.5!(i3)$);

        \node[left] (e1) at ($(i1) - (\fivesep,0)$){2};
        \node[left](e2) at ($(i3) - (\fivesep,0)$){1};
        
        \node[right] (e3) at ($(i1) + (\fivesep,0)$) {\(\ell_1\)};
        \node[right] (e4) at ($(cp) + (\fivesep,0)$) {\(\ell_2\)};
        \node[right](e5) at ($(i3) + (\fivesep,0)$){\(\ell_3\)};
        \draw (e1) -- (i1) -- (e3);
        \draw (e2) -- (i3) -- (e5);
        \draw (i2) -- (e4);
        \draw (i1) -- (i3);
      \end{tikzpicture}
      \\
      \begin{tikzpicture}
        \pgfmathsetmacro{\fivesep}{0.4}

        \coordinate(i1) at (0,\fivesep);
        \coordinate(i2) at (0.3*\fivesep,-\fivesep);
        \coordinate(i3) at (0,-\fivesep);
        \coordinate(cp) at ($(i1)!0.5!(i3)$);
        
        \node[left] (e1) at ($(i1) - (\fivesep,0)$){2};
        \node[left](e2) at ($(i3) - (\fivesep,0)$){1};
        
        \node[right] (e3) at ($(i1) + (\fivesep,0)$) {\(\ell_1\)};
        \node[right] (e4) at ($(cp) + (\fivesep,0)$){\(\ell_2\)};
        \node[right](e5) at ($(i3) + (\fivesep,0)$){\(\ell_3\)};
        \draw (e1) -- (i1) -- (e3);
        \draw (e2) -- (i3) -- (e5);
        \draw (i2) -- (e4);
        \draw (i1) -- (i3);
      \end{tikzpicture}
      \\
      \begin{tikzpicture}
       \pgfmathsetmacro{\fivesep}{0.4}

        \coordinate(i1) at (-\fivesep,0);
        \coordinate(i2) at (0,0);
        \coordinate(i3) at (\fivesep,0);
        \coordinate(cp) at ($(i1)!0.5!(i3)$);
        
        \node[left] (e1) at ($(i1) + (135:\fivesep)$){2};
        \node[left](e2) at ($(i1) + (225:\fivesep)$){1};
        
        \node[right] (e3) at ($(i3) + (45:\fivesep)$) {\(\ell_1\)};
        \node[right] (e4) at ($(i3) + (0:\fivesep)$){\(\ell_2\)};
        \node[right](e5) at ($(i3) + (-45:\fivesep)$){\(\ell_3\)};
        \draw (e1) -- (i1) -- (e2);
        \draw (cp) -- (e3);
        \draw (i3) -- (e4);
        \draw (i3) -- (e5);
        \draw (i1) -- (i3);
      \end{tikzpicture}
      \\
      \begin{tikzpicture}
        \pgfmathsetmacro{\fivesep}{0.4}

        \coordinate(i1) at (-\fivesep,0);
        \coordinate(i2) at (0,0);
        \coordinate(i3) at (\fivesep,0);
        \coordinate(cp) at ($(i1)!0.5!(i3)$);
        
        \node[left] (e1) at ($(i1) + (135:\fivesep)$){2};
        \node[left](e2) at ($(i1) + (225:\fivesep)$){1};
        
        \node[right] (e3) at ($(i3) + (45:\fivesep)$) {\(\ell_1\)};
        \node[right] (e4) at ($(i3) + (0:\fivesep)$){\(\ell_2\)};
        \node[right](e5) at ($(i3) + (-45:\fivesep)$){\(\ell_3\)};
        \draw (e1) -- (i1) -- (e2);
        \draw (cp) -- (e5);
        \draw (i3) -- (e4);
        \draw (i3) -- (e3);
        \draw (i1) -- (i3);
      \end{tikzpicture}
          \end{array}\right)
        \otimes
          \left(
    \begin{array}{c}
      \begin{tikzpicture}
        \pgfmathsetmacro{\fivesep}{0.4}

        \coordinate(i1) at (0,\fivesep);
        \coordinate(i2) at (0,0);
        \coordinate(i3) at (0,-\fivesep);
        \coordinate(cp) at ($(i1)!0.5!(i3)$);
        
        \node[right] (e1) at ($(i1) + (\fivesep,0)$){3};
        \node[right](e2) at ($(i3) + (\fivesep,0)$){4};
        
        \node[left] (e3) at ($(i1) - (\fivesep,0)$) {\(-\ell_1\)};
        \node[left] (e4) at ($(cp) - (\fivesep,0)$){\(-\ell_2\)};
        \node[left](e5) at ($(i3) - (\fivesep,0)$){\(-\ell_3\)};
        \draw (e1) -- (i1) -- (e3);
        \draw (e2) -- (i3) -- (e5);
        \draw (i2) -- (e4);
        \draw (i1) -- (i3);
      \end{tikzpicture}
      \\
      \begin{tikzpicture}
        \pgfmathsetmacro{\fivesep}{0.4}
        
        \coordinate(i1) at (0,\fivesep);
        \coordinate(i2) at (-0.3*\fivesep,\fivesep);
        \coordinate(i3) at (0,-\fivesep);
        \coordinate(cp) at ($(i1)!0.5!(i3)$);

        \node[right] (e1) at ($(i1) + (\fivesep,0)$){3};
        \node[right](e2) at ($(i3) + (\fivesep,0)$){4};
        
        \node[left] (e3) at ($(i1) - (\fivesep,0)$) {\(-\ell_1\)};
        \node[left] (e4) at ($(cp) - (\fivesep,0)$) {\(-\ell_2\)};
        \node[left](e5) at ($(i3) - (\fivesep,0)$){\(-\ell_3\)};
        \draw (e1) -- (i1) -- (e3);
        \draw (e2) -- (i3) -- (e5);
        \draw (i2) -- (e4);
        \draw (i1) -- (i3);
      \end{tikzpicture}
      \\
      \begin{tikzpicture}
        \pgfmathsetmacro{\fivesep}{0.4}

        \coordinate(i1) at (0,\fivesep);
        \coordinate(i2) at (-0.3*\fivesep,-\fivesep);
        \coordinate(i3) at (0,-\fivesep);
        \coordinate(cp) at ($(i1)!0.5!(i3)$);
        
        \node[right] (e1) at ($(i1) + (\fivesep,0)$){3};
        \node[right](e2) at ($(i3) + (\fivesep,0)$){4};
        
        \node[left] (e3) at ($(i1) - (\fivesep,0)$) {\(-\ell_1\)};
        \node[left] (e4) at ($(cp) - (\fivesep,0)$){\(-\ell_2\)};
        \node[left](e5) at ($(i3) - (\fivesep,0)$){\(-\ell_3\)};
        \draw (e1) -- (i1) -- (e3);
        \draw (e2) -- (i3) -- (e5);
        \draw (i2) -- (e4);
        \draw (i1) -- (i3);
      \end{tikzpicture}
      \\
      \begin{tikzpicture}
       \pgfmathsetmacro{\fivesep}{0.4}

        \coordinate(i1) at (\fivesep,0);
        \coordinate(i2) at (0,0);
        \coordinate(i3) at (-\fivesep,0);
        \coordinate(cp) at ($(i1)!0.5!(i3)$);
        
        \node[right] (e1) at ($(i1) + (45:\fivesep)$){3};
        \node[right](e2) at ($(i1) + (-45:\fivesep)$){4};
        
        \node[left] (e3) at ($(i3) + (135:\fivesep)$) {\(-\ell_1\)};
        \node[left] (e4) at ($(i3) + (180:\fivesep)$){\(-\ell_2\)};
        \node[left](e5) at ($(i3) + (225:\fivesep)$){\(-\ell_3\)};
        \draw (e1) -- (i1) -- (e2);
        \draw (cp) -- (e3);
        \draw (i3) -- (e4);
        \draw (i3) -- (e5);
        \draw (i1) -- (i3);
      \end{tikzpicture}
      \\
      \begin{tikzpicture}
        \pgfmathsetmacro{\fivesep}{0.4}

        \coordinate(i1) at (\fivesep,0);
        \coordinate(i2) at (0,0);
        \coordinate(i3) at (-\fivesep,0);
        \coordinate(cp) at ($(i1)!0.5!(i3)$);
        
        \node[right] (e1) at ($(i1) + (45:\fivesep)$){3};
        \node[right](e2) at ($(i1) + (-45:\fivesep)$){4};
        
        \node[left] (e3) at ($(i3) + (135:\fivesep)$) {\(-\ell_1\)};
        \node[left] (e4) at ($(i3) + (180:\fivesep)$){\(-\ell_2\)};
        \node[left](e5) at ($(i3) + (225:\fivesep)$){\(-\ell_3\)};
        \draw (e1) -- (i1) -- (e2);
        \draw (cp) -- (e5);
        \draw (i3) -- (e4);
        \draw (i3) -- (e3);
        \draw (i1) -- (i3);
      \end{tikzpicture}
          \end{array}\right)
\end{equation*}
\caption{A natural planar ordering for the 2-loop 3-particle cut
  topology is given by
  $A(1,2,\ell_1,\ell_2,\ell_3) \otimes
  A(-\ell_3,-\ell_2,-\ell_1,3,4)$.}
\label{fig:mu-cut2loops}
\end{figure}
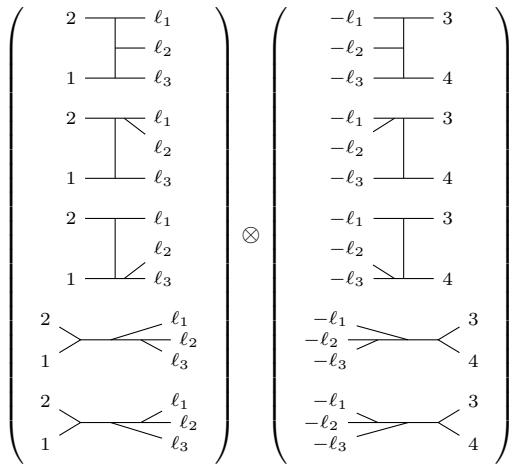
Finally, we turn our attention to the multi-unitarity cut topology,
\cref{fig:two-muc}, between two five-point trees.  A minimal $(n-3)!$
basis for the cut topology requires $((5-3)!)^2= 4$ orderings.  We can
find one planar ordering via the natural planar pairing between the
two amplitudes as depicted in \cref{fig:mu-cut2loops}.  Then, by
jointly permuting the ordering of the three internal legs, we find
$3! =6$ distinct color orderings which are topologically identical to
each other.  Any 4 of these 6 planar channels can be chosen as the
basis elements.  Even though it is not necessary for this cut
topology, it is worth pointing out that the cut topology actually
contains 12 distinct cut orderings.  The extra factor of 2 comes from
the ability to exchange \emph{one} of the pairs of external legs while
still preserving planarity,
\begin{align}
  &A(1,2,\ell_1,\ell_2,\ell_3) \otimes
  A(-\ell_3,-\ell_2,-\ell_1,3,4)  \notag \\
 \to &A(1,2,\ell_1,\ell_2,\ell_3) \otimes
       A(-\ell_3,-\ell_2,-\ell_1,4,3) \label{eq:ext-rev}\\
  \equiv &A(1,2,\ell_1,\ell_2,\ell_3) \otimes
           A(-\ell_1,-\ell_2,-\ell_3,3,4)\,. \notag
\end{align}
This reversal produces a distinct set of cut orderings because it is
equivalent to reversing the loop ordering on only one side of the cut,
\begin{equation}
A(1,2,\sigma) \otimes A(-\sigma^T,3,4)
    \neq
    A(1,2,\sigma) \otimes A(-\sigma,3,4)
    \label{eq:ext-rev-gen}
\end{equation}
Note that reversing both pairs of external legs is equivalent to
reversing the internal legs, and is thus already counted in the initial
set of permutations.

\begin{figure}[ht]
    \subfloat[The iterated unitarity cut]{
    \begin{tikzpicture}
      \coordinate (a1) at (0,0);
      \coordinate (a2) at (0.8,0);
      \coordinate (a3) at (1.6,0);
      \coordinate (a4) at (2.4,0);
      \draw pic {muc int={2}{a1}{a2}};
      \draw pic {muc int={2}{a2}{a3}};
      \draw pic {muc int={2}{a3}{a4}};
      \draw pic {exts={a1}{a4}};
      \node[blob] at (a1) {};
      \node[blob] at (a2) {};
      \node[blob] at (a3) {};
      \node[blob] at (a4) {};
      \draw pic {muc strut};
    \end{tikzpicture}
    \label{fig:three-unit}
  }\hfill
 \subfloat[The mixed iterated cut]{
    \begin{tikzpicture}
      \coordinate (a1) at (0,0);
      \coordinate (a2) at (1,0);
      \coordinate (a3) at (2,0);
      \draw pic {muc int={3}{a1}{a2}};
      \draw pic {muc int={2}{a2}{a3}};
      \draw pic {exts={a1}{a3}};
      \node[blob] at (a1) {};
      \node[blob] at (a2) {};
      \node[blob] at (a3) {};
      \draw pic {muc strut};
    \end{tikzpicture}
    \label{fig:three-mixed}
  }\\
 \subfloat[The first triangle unitarity cut]{
    \begin{tikzpicture}
      \coordinate (a1) at (0,0);
      \coordinate (a2) at (0,1);
      \coordinate (a3) at (1,0.5);
      \draw pic {muc int={3}{a1}{a2}};
      \draw pic {extr={a3}};
      \draw (a2) -- ++(135:0.7);
      \draw (a1) -- ++(225:0.7);
      \draw pic {cut line={a2}{a3}{0.3}};
      \draw pic {cut line={a1}{a3}{0.3}};
      \node[blob] at (a1) {};
      \node[blob] at (a2) {};
      \node[blob] at (a3) {};
    \end{tikzpicture}
    \label{fig:three-tri1}
  }\hfill
 \subfloat[The second triangle unitarity cut]{
    \begin{tikzpicture}
      \coordinate (a1) at (0,0);
      \coordinate (a2) at (0,1);
      \coordinate (a3) at (1,0.5);
      \draw pic {muc int scaled={2}{a1}{a2}{0.4}};
      \draw pic {muc int scaled={2}{a1}{a3}{0.3}};
      \draw pic {extr={a3}};
      \draw (a2) -- ++(135:0.7);
      \draw (a1) -- ++(225:0.7);
      \draw pic {cut line={a2}{a3}{0.3}};
      \node[blob] at (a1) {};
      \node[blob] at (a2) {};
      \node[blob] at (a3) {};
    \end{tikzpicture}
    \label{fig:three-tri2}
  }\hfill
 \subfloat[The multi-unitarity cut]{
   \begin{tikzpicture}
     \coordinate (s1) at (0,-0.5);
     \coordinate (s2) at (0,0.5);
     \path (s2) -- ++(135:0.7);
      \path (s1) -- ++(225:0.7);
     \draw pic {muc={4}};
    \end{tikzpicture}
    \label{fig:three-muc}
  }
\caption{The spanning physical generalized unitarity cuts for 3-loop 4-point.}
\label{fig:three-spanning}
\end{figure}
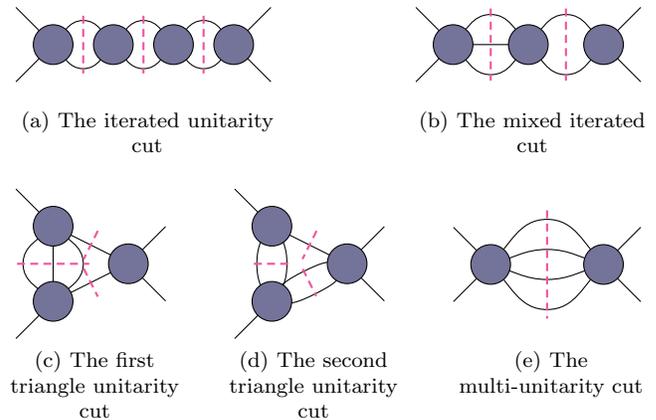

We proceed to three loops, where the spanning unitarity cuts are shown
in \cref{fig:three-spanning}.  As pointed out above, the iterated
unitarity cut topology, \cref{fig:three-unit}, is trivially planar
constructible.  The mixed iterated cut topology,
\cref{fig:three-mixed}, and the first of the generalized triangle cut
topologies, \cref{fig:three-tri1}, both contain a two-loop
multi-unitarity cut as a sub-graph.  Since the remainder of each graph
is simply a four-point amplitude, both cuts in their entirety are
planar constructible.  The second generalized triangle cut topology,
\cref{fig:three-tri2}, requires slightly more work.  This topology
contains two five-point and one four-point amplitudes, requiring 4
distinct orderings to span the minimal basis.  Unfortunately, we
cannot use the independent exchanges of both bubbles to provide all
necessary configurations, because the resulting orderings of the
bottom-left 5-point amplitude are not fully independent under the
Kleiss-Kuijf relations \cite{Kleiss:1988ne}.  However, we can use the
exchange of external legs of the right-hand five-point tree to provide
its two necessary contributions, while the bottom-left tree receives
two different orderings from the left-hand bubble. Thus we find
exactly the four orderings needed to span the minimal basis.

Finally, for the 3-loop multi-unitarity cut topology,
\cref{fig:three-muc}, we require $((6-3)!)^2= 36$ planar orderings to
span the basis, while the permutation of the loop legs produces
$4! = 24$ topologically identical orderings.  We see that here we
actually need the additional factor of two that comes from an external
reversal, as in \cref{eq:ext-rev,eq:ext-rev-gen}.  Thus we find
$2 \times 4! = 48$ fully planar orderings, which more than covers the
required $36$ to form the minimal basis.

Continuing the analysis to four loops, we see that all of the iterated
cuts can be constructed from planar data in analogy with three loops.
However, the counting for the multi-unitarity cut no longer supports a
manifestly planar solution, with $((7-3)!)^2= 576$ required basis
elements but only $2 \times 5! = 240$ planar orderings.  While this
removes the possibility of a general solution, we can restrict our
attention to maximal super-Yang--Mills, where the improved power
counting may admit a solution.  The expected power counting of the
theory means that, at worst, we expect the final new contact
contributions to come from N$^2$-max cuts.  We find that it is indeed
possible to find a planar-spanned basis for the N$^2$-max cuts of
4-loop 4-point maximal SYM (but, surprisingly, not the N$^3$-max
cuts), which we will elaborate on in a future work.

Given the above considerations, we can formulate the following general
statement for constructing four-point loop amplitudes using only
planar input:
\begin{equation}
  \begin{array}{ll}
    2 (L+1)! \ge (L!)^2 &\Rightarrow \text{Solution for all theories} \\
    2 (L+1)! < (L!)^2 &\Rightarrow \text{Only special cases}
  \end{array}
\end{equation}
where the crossover point for the constraint is between $L=3$ and
$L=4$.

\section{Conclusions}
In this letter, we have introduced the method of color-dual cut
tiling, which drastically reduces the number of ordered cuts needed to
construct non-color-dual gauge theory integrands. We apply this method
to the four-point amplitude of the $\mathcal{N}=4$ SYM theory at
three-loops to show that the entire integrand can be reconstructed
from knowledge of only the planar sector.  Intriguingly, the 3-loop
4-point $\mathcal{N}=8$ supergravity integrand is therefore
\emph{entirely} constrained by the positivity properties of the planar
$\mathcal{N}=4$ SYM theory via color-dual cut tiling and generalized
double-copy.  This provides further~\cite{Bern:2015ple,
  Edison:2019ovj} evidence of combinatoric geometric structures which
encode the dynamics of non-planar theories.  Furthermore, we have
demonstrated that the two- and three-loop four-point amplitudes of
\emph{any} tree-level color-dual theory can be constructed using only
planar diagram numerators as input.

While our combinatoric arguments no longer guarantee success at four
loops, we find that the improved power-counting of the $\mathcal{N}=4$
SYM theory sufficiently limits the number of relevant cuts such that
all of them can also be constructed in terms of only planar data.  It
will be interesting to see how the improved power-counting affects the
viability of our process at higher loops and more legs.

We close by commenting that evaluating the UV behavior of
$\mathcal{N}=5$ SG at five loops and $\mathcal{N}=8$ SG at seven loops
poses a challenge that clearly requires innovation in terms of
integrand construction.  For calculations where explicit cut
evaluation becomes expensive, we expect color-dual cut tiling can help
in conjunction with recent innovations in dressing integrands
efficiently~\cite{Carrasco:2021otn,Bern:2024vqs}.

\section*{Acknowledgments} 
We thank Sasank Chava, Henrik Johansson, Eliseu Kloster, Shruti
Paranjape, Nic Pavao, Aslan Seifi, Cong Shen, and Jaroslav Trnka for
inspiring conversations.
NR is grateful for the support of a grant
from the Undergraduate Research Grant Program which is administered by
Northwestern University's Office of Undergraduate Research thanks to
the generous donation of the Hung-Farineli family.
SZ is grateful for the support from Northwestern University’s
Presidential Fellowship.
This work was supported by the DOE under contract DE-SC0015910, by the
Alfred P. Sloan Foundation, and by Northwestern University via the
Amplitudes and Insight Group, Department of Physics and Astronomy, and
Weinberg College of Arts and Sciences.

This research was supported in part through the computational
resources and staff contributions provided for the Quest high
performance computing facility at Northwestern University which is
jointly supported by the Office of the Provost, the Office for
Research, and Northwestern University Information Technology.

fill\TeX ~was used as part of writing the bibliography
\cite{2017JOSS....2..222G}.

\bibliography{letter-bib}

\end{document}

%% file: tikz_preamble.tex
\usetikzlibrary{calc}
\usetikzlibrary{math}
\usetikzlibrary{positioning}
\usetikzlibrary{decorations.pathreplacing,calligraphy}
\usetikzlibrary {arrows.meta}

\makeatletter
\def\convertto#1#2{\strip@pt\dimexpr #2*65536/\number\dimexpr 1#1}
\makeatother

\newcommand{\pgfextractangle}[3]{%
  \pgfmathanglebetweenpoints{\pgfpointanchor{#2}{center}}
  {\pgfpointanchor{#3}{center}}
  \global\let#1\pgfmathresult  
}

\pgfmathsetmacro{\extL}{0.7}

\tikzset{every node/.style={font=\scriptsize}}

\tikzset{cuts/.style={dash pattern=on 3pt off 2pt,
    draw=VioletRed,
    thick
  }}

\tikzset{np/.style={draw=white,line width=3pt}}

\tikzset{blob/.style={draw=black,fill=blobcolor,circle,minimum size=15,inner sep=0}}

\tikzset{small blob/.style={blob,minimum size =8}}

\tikzset{pics/cut line/.style n args={3}{code={%
      \draw (#1) -- (#2);
      \pgfextractangle{\orient}{#1}{#2}
      \coordinate (mp) at ($(#1)!0.5!(#2)$);
      \draw[cuts] ($(mp) + (90+\orient:1.4*#3/2)$) -- ($(mp) + (-90+\orient:1.4*#3/2)$);
    }
  }
}

\tikzset{noncut/.style ={thick}}

\tikzset{pics/muc int scaled/.style n args={4}{code={%
      
      \coordinate (mp) at ($(#2)!0.5!(#3)$);
      \tikzmath{coordinate \C;
        \C = (#3)-(#2);
        \distMUC = sqrt((\Cx)^2+(\Cy)^2);
        \distMUC = \convertto{cm}{\distMUC pt}; 
      }
      \pgfmathsetmacro{\mh}{#4*\distMUC};
      
      \pgfextractangle{\orient}{#2}{#3}
      
      \pgfmathsetmacro{\theta}{(90/(#1-1))};
      \pgfmathsetmacro{\midsep}{(\mh/(#1-1))};
      \pgfmathsetmacro{\forend}{(#1-1)}
      \foreach \ml in {0,...,\forend}
      {
        \pgfmathsetmacro\thetao{\orient+45-\theta*\ml};
        \pgfmathsetmacro\thetai{\orient+135+\theta*\ml};
        \coordinate (\ml) at ($(mp) + (90+\orient:\mh/2-\midsep*\ml)$);
        \draw (#2) to[out=\thetao,in=180+\orient] (\ml)
        to[out=\orient,in=\thetai] (#3);
      }
      \draw[cuts] ($(mp) + (90+\orient:1.2*\mh/2)$) -- ($(mp) + (-90+\orient:1.2*\mh/2)$);
    }
  }
}

\tikzset{pics/muc int/.style n args={3}{code={
      \draw pic {muc int scaled={#1}{#2}{#3}{0.8}};
    }
  }
}

\tikzset{pics/exts/.style n args={2}{code={%
      \draw (#1) -- ++(135:\extL);
      \draw (#1) -- ++(225:\extL);

      \draw (#2) -- ++(45:\extL);
      \draw (#2) -- ++(-45:\extL);
    }
  }
}

\tikzset{pics/extl/.style n args={1}{code={%
      \draw (#1) -- ++(135:\extL);
      \draw (#1) -- ++(225:\extL);
    }
  }
}

\tikzset{pics/extr/.style n args={1}{code={%
      \draw (#1) -- ++(45:\extL);
      \draw (#1) -- ++(-45:\extL);
    }
  }
}

\tikzset{pics/muc strut/.style={code={%
      \coordinate(a1) at (0,0);
      \coordinate(a2) at (1.5,0);
      \tikzmath{coordinate \C;
        \C = (a2)-(a1);
        \distMUC = sqrt((\Cx)^2+(\Cy)^2);
        \distMUC = \convertto{cm}{\distMUC pt}; 
      }
      \pgfmathsetmacro{\mh}{0.8*\distMUC};
      \path (0,1.2*\mh/2) -- (0,-1.2*\mh/2);
    }
  }
}

\tikzset{pics/muc/.style n args={1}{code={%
      \coordinate(a1) at (0,0);
      \coordinate(a2) at (1.5,0);

      \draw pic {exts={a1}{a2}};
      
      \draw pic {muc int={#1}{a1}{a2}};
      \node[blob] at (a1) {};
      \node[blob] at (a2) {};
    }
  }
}

\tikzset{pics/tri skel/.style n args={1}{code={%
      \coordinate (vl) at (180:#1);
      \coordinate (vu) at (60:#1);
      \coordinate (vd) at (-60:#1);
      \coordinate (mid) at (0,0);
      \coordinate (mu) at ($(vl)!0.5!(vu)$);
      \coordinate (mr) at ($(vu)!0.5!(vd)$);
      \coordinate (ml) at ($(vd)!0.5!(vl)$);
    }
  }
}

\tikzset{pics/tri skel2/.style n args={1}{code={%
      \coordinate (vl) at (180:#1);
      \coordinate (vu) at (60:#1);
      \coordinate (vd) at (-60:#1);
      \coordinate (mid) at (0,0);
      \coordinate (mu1) at ($(vl)!0.33!(vu)$);
      \coordinate (mu2) at ($(vl)!0.66!(vu)$);
      \coordinate (mr1) at ($(vu)!0.33!(vd)$);
      \coordinate (mr2) at ($(vu)!0.66!(vd)$);
      \coordinate (ml1) at ($(vd)!0.33!(vl)$);
      \coordinate (ml2) at ($(vd)!0.66!(vl)$);
    }
  }
}

\tikzset{pics/tri skel gen/.style n args={4}{code={%
      \coordinate (vl) at (180:#1);
      \coordinate (vu) at (60:#1);
      \coordinate (vd) at (-60:#1);
      \coordinate (mid) at (0,0);
      \pgfmathsetmacro{\prop}{1.0/(#2+1)};
      \foreach \id in {1,...,#2}{
        \coordinate (mu\id) at ($(vl)!\id*\prop!(vu)$);
      }
      \pgfmathsetmacro{\prop}{1.0/(#3+1)};
      \foreach \id in {1,...,#3}{
        \coordinate (mr\id) at ($(vu)!\id*\prop!(vd)$);
      }
      \pgfmathsetmacro{\prop}{1.0/(#4+1)};
      \foreach \id in {1,...,#4}{
        \coordinate (md\id) at ($(vd)!\id*\prop!(vl)$);
      }
    }
  }
}

\tikzset{pics/box skel gen/.style n args={5}{code={%
      \coordinate (vlu) at (135:#1);
      \coordinate (vld) at (-135:#1);
      \coordinate (vru) at (45:#1);
      \coordinate (vrd) at (-45:#1);
      \coordinate (mid) at (0,0);
      \pgfmathsetmacro{\prop}{1.0/(#2+1)};
      \foreach \id in {1,...,#2}{
        \coordinate (mu\id) at ($(vlu)!\id*\prop!(vru)$);
      }
      \pgfmathsetmacro{\prop}{1.0/(#3+1)};
      \foreach \id in {1,...,#3}{
        \coordinate (mr\id) at ($(vru)!\id*\prop!(vrd)$);
      }
      \pgfmathsetmacro{\prop}{1.0/(#4+1)};
      \foreach \id in {1,...,#4}{
        \coordinate (md\id) at ($(vrd)!\id*\prop!(vld)$);
      }
      \pgfmathsetmacro{\prop}{1.0/(#5+1)};
      \foreach \id in {1,...,#5}{
        \coordinate (ml\id) at ($(vld)!\id*\prop!(vlu)$);
      }
    }
  }
}

\tikzset{pics/box skel/.style n args={1}{code={%
      \coordinate (vlu) at (135:#1);
      \coordinate (vld) at (-135:#1);
      \coordinate (vru) at (45:#1);
      \coordinate (vrd) at (-45:#1);
      \coordinate (mid) at (0,0);
      \coordinate (mu) at ($(vlu)!0.5!(vru)$);
      \coordinate (md) at ($(vld)!0.5!(vrd)$);
      \coordinate (mr) at ($(vru)!0.5!(vrd)$);
      \coordinate (ml) at ($(vld)!0.5!(vlu)$);
    }
  }
}

\tikzset{pics/diag a/.style n args={1}{code={%
      \pic{box skel gen={#1}{2}{0}{2}{0}};
      \draw[noncut] (vru) -- ++(45:\extL/2);
      \draw[noncut]  (vrd) -- ++(-45:\extL/2);
      \draw[noncut]  (vlu) -- ++(135:\extL/2);
      \draw[noncut]  (vld) -- ++(225:\extL/2);
      \draw[noncut]  (mu1) -- (md2);
      \draw[noncut] (mu2) -- (md1);
      
      \draw[noncut] (vlu) -- (vru) -- (vrd) -- (vld) -- (vlu);
      \node[] at (270:#1){(a)};
    }
  }
}

\tikzset{pics/diag b/.style n args={1}{code={%
      \pic{box skel gen={#1}{2}{0}{2}{0}};
      \draw[noncut](vru) -- ++(45:\extL/2);
      \draw[noncut] (vrd) -- ++(-45:\extL/2);
      \draw[noncut] (vlu) -- ++(135:\extL/2);
      \draw[noncut] (vld) -- ++(225:\extL/2);
      
      \draw[noncut] (mu1) -- (md2);
      \draw[noncut](mu2) -- (vrd);

      \draw[np] (md1)--(vru);
      \draw[noncut] (md1)--(vru);
      
      \draw[noncut] (vru) --(vlu) -- (vld) -- (vrd);
      \node[] at (270:#1){(b)};
    }
  }
}

\tikzset{pics/diag c/.style n args={1}{code={%
      \pic{box skel gen={#1}{2}{0}{2}{0}};
      \draw[noncut] (vru) -- ++(45:\extL/2);
      \draw[noncut] (vrd) -- ++(-45:\extL/2);
      \draw[noncut] (vlu) -- ++(135:\extL/2);
      \draw[noncut] (vld) -- ++(225:\extL/2);
      \draw[noncut] (mu1) -- (md1);
      
      \draw[np] (mu2) -- (md2);
      \draw[noncut] (mu2) -- (md2);
      
      \draw[noncut] (vlu) -- (vru) -- (vrd) -- (vld) -- (vlu);
      \node[] at (270:#1){(c)};
    }
  }
}
\tikzset{pics/diag d/.style n args={1}{code={%
      \pic{box skel gen={#1}{2}{0}{2}{0}};
      \draw[noncut] (vru) -- ++(45:\extL/2);
      \draw[noncut] (vrd) -- ++(-45:\extL/2);
      \draw[noncut] (vlu) -- ++(135:\extL/2);
      \draw[noncut] (vld) -- ++(225:\extL/2);
      
      \draw[noncut] (md2) -- (vlu);
      \draw[np] (vld) -- (mu1);
      \draw[noncut] (vld) -- (mu1);

      \draw[noncut] (mu2) -- (vrd);
      \draw[np] (md1)--(vru);
      \draw[noncut] (md1)--(vru);
      
      \draw[noncut] (vru) --(vlu);
      \draw[noncut] (vld) -- (vrd);
      \node[] at (270:#1){(d)};
    }
  }
}

\tikzset{pics/diag e/.style n args={1}{code={%
      \pic{box skel gen={#1}{1}{1}{1}{0}};
      \draw[noncut] (vru) -- ++(45:\extL/2);
      \draw[noncut] (vrd) -- ++(-45:\extL/2);
      \draw[noncut] (vlu) -- ++(135:\extL/2);
      \draw[noncut] (vld) -- ++(225:\extL/2);
      \draw[noncut] (mid) -- (mr1);
      \draw[noncut] (mu1) -- (md1);
      \draw[noncut] (vlu) -- (vru) -- (vrd) -- (vld) -- (vlu);
      \node[] at (270:#1){(e)};
    }
  }
}

\tikzset{pics/diag f/.style n args={1}{code={%
      \pic{box skel gen={#1}{1}{1}{1}{0}};
      \draw[noncut] (vru) -- ++(45:\extL/2);
      \draw[noncut] (vrd) -- ++(-45:\extL/2);
      \draw[noncut] (vlu) -- ++(135:\extL/2);
      \draw[noncut] (vld) -- ++(225:\extL/2);
   
      \draw[noncut] (mu1) -- (mr1);
      \draw[np](mid) -- (vru);
      \draw[noncut](mid) -- (vru);
      \draw[noncut] (md1)--(mid) -- (mr1)--(vrd);
        
      \draw[noncut] (vru)-- (vlu) --  (vld) -- (vrd);
      \node[] at (270:#1){(f)};
    }
  }
}

\tikzset{pics/diag g/.style n args={1}{code={%
      \pic{box skel gen={#1}{1}{1}{1}{0}};
      \draw[noncut] (vru) -- ++(45:\extL/2);
      \draw[noncut] (vrd) -- ++(-45:\extL/2);
      \draw[noncut] (vlu) -- ++(135:\extL/2);
      \draw[noncut] (vld) -- ++(225:\extL/2);
   
      \draw[noncut] (mu1) -- (mr1);
      \draw[np](mid) -- (vru);
      \draw[noncut](mid) -- (vru);
      \draw[noncut] (md1)--(mid) --(mu1);
      \draw[noncut] (vrd)--(vru);
        
      \draw[noncut] (mu1)-- (vlu) --  (vld) -- (vrd);
      \node[] at (270:#1){(g)};
    }
  }
}

\tikzset{pics/diag h/.style n args={1}{code={%
      \pic{box skel gen={#1}{1}{1}{1}{1}};
      \draw[noncut] (vru) -- ++(45:\extL/2);
      \draw[noncut] (vrd) -- ++(-45:\extL/2);
      \draw[noncut] (vlu) -- ++(135:\extL/2);
      \draw[noncut] (vld) -- ++(225:\extL/2);
   
      \draw[noncut](mr1) -- (ml1);
      \draw[np](md1) -- (mu1);
      \draw[noncut](md1) -- (mu1);
        
      \draw[noncut] (vlu) -- (vru) -- (vrd) -- (vld) -- (vlu);
      \node[] at (270:#1){(h)};
    }
  }
}

\tikzset{pics/diag i/.style n args={1}{code={%
      \pic{box skel gen={#1}{1}{1}{1}{0}};
      \draw[noncut] (vru) -- ++(45:\extL/2);
      \draw[noncut] (vrd) -- ++(-45:\extL/2);
      \draw[noncut] (vlu) -- ++(135:\extL/2);
      \draw[noncut] (vld) -- ++(225:\extL/2);
   
      \draw[noncut](vld) -- (mu1);

      \draw[np](mid) -- (vlu);
      \draw[noncut](mid) -- (vlu);
      \draw[noncut](md1)--(mid)--(mr1);

      \draw[noncut] (vlu) -- (vru) -- (vrd) -- (vld);
      \node[] at (270:#1){(i)};
    }
  }
}

%% file: letter.bbl
\begin{thebibliography}{43}%
\makeatletter
\providecommand \@ifxundefined [1]{%
 \@ifx{#1\undefined}
}%
\providecommand \@ifnum [1]{%
 \ifnum #1\expandafter \@firstoftwo
 \else \expandafter \@secondoftwo
 \fi
}%
\providecommand \@ifx [1]{%
 \ifx #1\expandafter \@firstoftwo
 \else \expandafter \@secondoftwo
 \fi
}%
\providecommand \natexlab [1]{#1}%
\providecommand \enquote  [1]{``#1''}%
\providecommand \bibnamefont  [1]{#1}%
\providecommand \bibfnamefont [1]{#1}%
\providecommand \citenamefont [1]{#1}%
\providecommand \href@noop [0]{\@secondoftwo}%
\providecommand \href [0]{\begingroup \@sanitize@url \@href}%
\providecommand \@href[1]{\@@startlink{#1}\@@href}%
\providecommand \@@href[1]{\endgroup#1\@@endlink}%
\providecommand \@sanitize@url [0]{\catcode `\\12\catcode `\$12\catcode
  `\&12\catcode `\#12\catcode `\^12\catcode `\_12\catcode `\%12\relax}%
\providecommand \@@startlink[1]{}%
\providecommand \@@endlink[0]{}%
\providecommand \url  [0]{\begingroup\@sanitize@url \@url }%
\providecommand \@url [1]{\endgroup\@href {#1}{\urlprefix }}%
\providecommand \urlprefix  [0]{URL }%
\providecommand \Eprint [0]{\href }%
\providecommand \doibase [0]{http://dx.doi.org/}%
\providecommand \selectlanguage [0]{\@gobble}%
\providecommand \bibinfo  [0]{\@secondoftwo}%
\providecommand \bibfield  [0]{\@secondoftwo}%
\providecommand \translation [1]{[#1]}%
\providecommand \BibitemOpen [0]{}%
\providecommand \bibitemStop [0]{}%
\providecommand \bibitemNoStop [0]{.\EOS\space}%
\providecommand \EOS [0]{\spacefactor3000\relax}%
\providecommand \BibitemShut  [1]{\csname bibitem#1\endcsname}%
\let\auto@bib@innerbib\@empty
\bibitem [{\citenamefont {Bern}\ \emph {et~al.}(1994)\citenamefont {Bern},
  \citenamefont {Dixon}, \citenamefont {Dunbar},\ and\ \citenamefont
  {Kosower}}]{Bern:1994zx}%
  \BibitemOpen
  \bibfield  {author} {\bibinfo {author} {\bibfnamefont {Zvi}\ \bibnamefont
  {Bern}}, \bibinfo {author} {\bibfnamefont {Lance~J.}\ \bibnamefont {Dixon}},
  \bibinfo {author} {\bibfnamefont {David~C.}\ \bibnamefont {Dunbar}}, \ and\
  \bibinfo {author} {\bibfnamefont {David~A.}\ \bibnamefont {Kosower}},\
  }\bibfield  {title} {\enquote {\bibinfo {title} {{One loop n point gauge
  theory amplitudes, unitarity and collinear limits}},}\ }\href {\doibase
  10.1016/0550-3213(94)90179-1} {\bibfield  {journal} {\bibinfo  {journal}
  {Nucl. Phys. B}\ }\textbf {\bibinfo {volume} {425}},\ \bibinfo {pages}
  {217--260} (\bibinfo {year} {1994})},\ \Eprint
  {http://arxiv.org/abs/hep-ph/9403226} {arXiv:hep-ph/9403226} \BibitemShut
  {NoStop}%
\bibitem [{\citenamefont {Bern}\ \emph {et~al.}(1995)\citenamefont {Bern},
  \citenamefont {Dixon}, \citenamefont {Dunbar},\ and\ \citenamefont
  {Kosower}}]{Bern:1994cg}%
  \BibitemOpen
  \bibfield  {author} {\bibinfo {author} {\bibfnamefont {Zvi}\ \bibnamefont
  {Bern}}, \bibinfo {author} {\bibfnamefont {Lance~J.}\ \bibnamefont {Dixon}},
  \bibinfo {author} {\bibfnamefont {David~C.}\ \bibnamefont {Dunbar}}, \ and\
  \bibinfo {author} {\bibfnamefont {David~A.}\ \bibnamefont {Kosower}},\
  }\bibfield  {title} {\enquote {\bibinfo {title} {{Fusing gauge theory tree
  amplitudes into loop amplitudes}},}\ }\href {\doibase
  10.1016/0550-3213(94)00488-Z} {\bibfield  {journal} {\bibinfo  {journal}
  {Nucl. Phys. B}\ }\textbf {\bibinfo {volume} {435}},\ \bibinfo {pages}
  {59--101} (\bibinfo {year} {1995})},\ \Eprint
  {http://arxiv.org/abs/hep-ph/9409265} {arXiv:hep-ph/9409265} \BibitemShut
  {NoStop}%
\bibitem [{\citenamefont {Britto}\ \emph {et~al.}(2005)\citenamefont {Britto},
  \citenamefont {Cachazo},\ and\ \citenamefont {Feng}}]{Britto:2004nc}%
  \BibitemOpen
  \bibfield  {author} {\bibinfo {author} {\bibfnamefont {Ruth}\ \bibnamefont
  {Britto}}, \bibinfo {author} {\bibfnamefont {Freddy}\ \bibnamefont
  {Cachazo}}, \ and\ \bibinfo {author} {\bibfnamefont {Bo}~\bibnamefont
  {Feng}},\ }\bibfield  {title} {\enquote {\bibinfo {title} {{Generalized
  unitarity and one-loop amplitudes in N=4 super-Yang-Mills}},}\ }\href
  {\doibase 10.1016/j.nuclphysb.2005.07.014} {\bibfield  {journal} {\bibinfo
  {journal} {Nucl. Phys. B}\ }\textbf {\bibinfo {volume} {725}},\ \bibinfo
  {pages} {275--305} (\bibinfo {year} {2005})},\ \Eprint
  {http://arxiv.org/abs/hep-th/0412103} {arXiv:hep-th/0412103} \BibitemShut
  {NoStop}%
\bibitem [{\citenamefont {Anastasiou}\ \emph {et~al.}(2007)\citenamefont
  {Anastasiou}, \citenamefont {Britto}, \citenamefont {Feng}, \citenamefont
  {Kunszt},\ and\ \citenamefont {Mastrolia}}]{Anastasiou:2006jv}%
  \BibitemOpen
  \bibfield  {author} {\bibinfo {author} {\bibfnamefont {Charalampos}\
  \bibnamefont {Anastasiou}}, \bibinfo {author} {\bibfnamefont {Ruth}\
  \bibnamefont {Britto}}, \bibinfo {author} {\bibfnamefont {Bo}~\bibnamefont
  {Feng}}, \bibinfo {author} {\bibfnamefont {Zoltan}\ \bibnamefont {Kunszt}}, \
  and\ \bibinfo {author} {\bibfnamefont {Pierpaolo}\ \bibnamefont
  {Mastrolia}},\ }\bibfield  {title} {\enquote {\bibinfo {title}
  {{D-dimensional unitarity cut method}},}\ }\href {\doibase
  10.1016/j.physletb.2006.12.022} {\bibfield  {journal} {\bibinfo  {journal}
  {Phys. Lett. B}\ }\textbf {\bibinfo {volume} {645}},\ \bibinfo {pages}
  {213--216} (\bibinfo {year} {2007})},\ \Eprint
  {http://arxiv.org/abs/hep-ph/0609191} {arXiv:hep-ph/0609191} \BibitemShut
  {NoStop}%
\bibitem [{\citenamefont {Bern}\ \emph {et~al.}(2007)\citenamefont {Bern},
  \citenamefont {Carrasco}, \citenamefont {Johansson},\ and\ \citenamefont
  {Kosower}}]{Bern:2007ct}%
  \BibitemOpen
  \bibfield  {author} {\bibinfo {author} {\bibfnamefont {Z.}~\bibnamefont
  {Bern}}, \bibinfo {author} {\bibfnamefont {J.~J.~M.}\ \bibnamefont
  {Carrasco}}, \bibinfo {author} {\bibfnamefont {Henrik}\ \bibnamefont
  {Johansson}}, \ and\ \bibinfo {author} {\bibfnamefont {D.~A.}\ \bibnamefont
  {Kosower}},\ }\bibfield  {title} {\enquote {\bibinfo {title} {{Maximally
  supersymmetric planar Yang-Mills amplitudes at five loops}},}\ }\href
  {\doibase 10.1103/PhysRevD.76.125020} {\bibfield  {journal} {\bibinfo
  {journal} {Phys. Rev. D}\ }\textbf {\bibinfo {volume} {76}},\ \bibinfo
  {pages} {125020} (\bibinfo {year} {2007})},\ \Eprint
  {http://arxiv.org/abs/0705.1864} {arXiv:0705.1864 [hep-th]} \BibitemShut
  {NoStop}%
\bibitem [{\citenamefont {Carrasco}\ and\ \citenamefont
  {Pavao}(2024)}]{Carrasco:2023qgz}%
  \BibitemOpen
  \bibfield  {author} {\bibinfo {author} {\bibfnamefont {John Joseph~M.}\
  \bibnamefont {Carrasco}}\ and\ \bibinfo {author} {\bibfnamefont {Nicolas~H.}\
  \bibnamefont {Pavao}},\ }\bibfield  {title} {\enquote {\bibinfo {title}
  {{Even-point multi-loop unitarity and its applications: exponentiation,
  anomalies and evanescence}},}\ }\href {\doibase 10.1007/JHEP01(2024)019}
  {\bibfield  {journal} {\bibinfo  {journal} {JHEP}\ }\textbf {\bibinfo
  {volume} {01}},\ \bibinfo {pages} {019} (\bibinfo {year} {2024})},\ \Eprint
  {http://arxiv.org/abs/2307.16812} {arXiv:2307.16812 [hep-th]} \BibitemShut
  {NoStop}%
\bibitem [{\citenamefont {Travaglini}\ \emph {et~al.}(2022)\citenamefont
  {Travaglini} \emph {et~al.}}]{Travaglini:2022uwo}%
  \BibitemOpen
  \bibfield  {author} {\bibinfo {author} {\bibfnamefont {Gabriele}\
  \bibnamefont {Travaglini}} \emph {et~al.},\ }\bibfield  {title} {\enquote
  {\bibinfo {title} {{The SAGEX review on scattering amplitudes}},}\ }\href
  {\doibase 10.1088/1751-8121/ac8380} {\bibfield  {journal} {\bibinfo
  {journal} {J. Phys. A}\ }\textbf {\bibinfo {volume} {55}},\ \bibinfo {pages}
  {443001} (\bibinfo {year} {2022})},\ \Eprint
  {http://arxiv.org/abs/2203.13011} {arXiv:2203.13011 [hep-th]} \BibitemShut
  {NoStop}%
\bibitem [{\citenamefont {Bern}\ \emph {et~al.}(2022)\citenamefont {Bern},
  \citenamefont {Carrasco}, \citenamefont {Chiodaroli}, \citenamefont
  {Johansson},\ and\ \citenamefont {Roiban}}]{Bern:2022wqg}%
  \BibitemOpen
  \bibfield  {author} {\bibinfo {author} {\bibfnamefont {Zvi}\ \bibnamefont
  {Bern}}, \bibinfo {author} {\bibfnamefont {John~Joseph}\ \bibnamefont
  {Carrasco}}, \bibinfo {author} {\bibfnamefont {Marco}\ \bibnamefont
  {Chiodaroli}}, \bibinfo {author} {\bibfnamefont {Henrik}\ \bibnamefont
  {Johansson}}, \ and\ \bibinfo {author} {\bibfnamefont {Radu}\ \bibnamefont
  {Roiban}},\ }\bibfield  {title} {\enquote {\bibinfo {title} {{The SAGEX
  review on scattering amplitudes Chapter 2: An invitation to color-kinematics
  duality and the double copy}},}\ }\href {\doibase 10.1088/1751-8121/ac93cf}
  {\bibfield  {journal} {\bibinfo  {journal} {J. Phys. A}\ }\textbf {\bibinfo
  {volume} {55}},\ \bibinfo {pages} {443003} (\bibinfo {year} {2022})},\
  \Eprint {http://arxiv.org/abs/2203.13013} {arXiv:2203.13013 [hep-th]}
  \BibitemShut {NoStop}%
\bibitem [{\citenamefont {Carrasco}\ \emph {et~al.}(2021)\citenamefont
  {Carrasco}, \citenamefont {Edison},\ and\ \citenamefont
  {Johansson}}]{Carrasco:2021otn}%
  \BibitemOpen
  \bibfield  {author} {\bibinfo {author} {\bibfnamefont {John Joseph~M.}\
  \bibnamefont {Carrasco}}, \bibinfo {author} {\bibfnamefont {Alex}\
  \bibnamefont {Edison}}, \ and\ \bibinfo {author} {\bibfnamefont {Henrik}\
  \bibnamefont {Johansson}},\ }\bibfield  {title} {\enquote {\bibinfo {title}
  {{Maximal Super-Yang-Mills at Six Loops via Novel Integrand Bootstrap}},}\
  }\href@noop {} {\  (\bibinfo {year} {2021})},\ \Eprint
  {http://arxiv.org/abs/2112.05178} {arXiv:2112.05178 [hep-th]} \BibitemShut
  {NoStop}%
\bibitem [{\citenamefont {Bern}\ \emph {et~al.}(2018)\citenamefont {Bern},
  \citenamefont {Carrasco}, \citenamefont {Chen}, \citenamefont {Edison},
  \citenamefont {Johansson}, \citenamefont {Parra-Martinez}, \citenamefont
  {Roiban},\ and\ \citenamefont {Zeng}}]{Bern:2018jmv}%
  \BibitemOpen
  \bibfield  {author} {\bibinfo {author} {\bibfnamefont {Zvi}\ \bibnamefont
  {Bern}}, \bibinfo {author} {\bibfnamefont {John~Joseph}\ \bibnamefont
  {Carrasco}}, \bibinfo {author} {\bibfnamefont {Wei-Ming}\ \bibnamefont
  {Chen}}, \bibinfo {author} {\bibfnamefont {Alex}\ \bibnamefont {Edison}},
  \bibinfo {author} {\bibfnamefont {Henrik}\ \bibnamefont {Johansson}},
  \bibinfo {author} {\bibfnamefont {Julio}\ \bibnamefont {Parra-Martinez}},
  \bibinfo {author} {\bibfnamefont {Radu}\ \bibnamefont {Roiban}}, \ and\
  \bibinfo {author} {\bibfnamefont {Mao}\ \bibnamefont {Zeng}},\ }\bibfield
  {title} {\enquote {\bibinfo {title} {{Ultraviolet Properties of $\mathcal N =
  8$ Supergravity at Five Loops}},}\ }\href {\doibase
  10.1103/PhysRevD.98.086021} {\bibfield  {journal} {\bibinfo  {journal} {Phys.
  Rev. D}\ }\textbf {\bibinfo {volume} {98}},\ \bibinfo {pages} {086021}
  (\bibinfo {year} {2018})},\ \Eprint {http://arxiv.org/abs/1804.09311}
  {arXiv:1804.09311 [hep-th]} \BibitemShut {NoStop}%
\bibitem [{\citenamefont {Bern}\ \emph {et~al.}(2012)\citenamefont {Bern},
  \citenamefont {Carrasco}, \citenamefont {Dixon}, \citenamefont {Johansson},\
  and\ \citenamefont {Roiban}}]{Bern:2012uf}%
  \BibitemOpen
  \bibfield  {author} {\bibinfo {author} {\bibfnamefont {Z.}~\bibnamefont
  {Bern}}, \bibinfo {author} {\bibfnamefont {J.~J.~M.}\ \bibnamefont
  {Carrasco}}, \bibinfo {author} {\bibfnamefont {L.~J.}\ \bibnamefont {Dixon}},
  \bibinfo {author} {\bibfnamefont {H.}~\bibnamefont {Johansson}}, \ and\
  \bibinfo {author} {\bibfnamefont {R.}~\bibnamefont {Roiban}},\ }\bibfield
  {title} {\enquote {\bibinfo {title} {{Simplifying Multiloop Integrands and
  Ultraviolet Divergences of Gauge Theory and Gravity Amplitudes}},}\ }\href
  {\doibase 10.1103/PhysRevD.85.105014} {\bibfield  {journal} {\bibinfo
  {journal} {Phys. Rev. D}\ }\textbf {\bibinfo {volume} {85}},\ \bibinfo
  {pages} {105014} (\bibinfo {year} {2012})},\ \Eprint
  {http://arxiv.org/abs/1201.5366} {arXiv:1201.5366 [hep-th]} \BibitemShut
  {NoStop}%
\bibitem [{\citenamefont {Edison}\ and\ \citenamefont
  {Tegevi}(2023)}]{Edison:2022smn}%
  \BibitemOpen
  \bibfield  {author} {\bibinfo {author} {\bibfnamefont {Alex}\ \bibnamefont
  {Edison}}\ and\ \bibinfo {author} {\bibfnamefont {Micah}\ \bibnamefont
  {Tegevi}},\ }\bibfield  {title} {\enquote {\bibinfo {title}
  {{Color-kinematics dual representations of one-loop matrix elements in the
  open-superstring effective action}},}\ }\href {\doibase
  10.1007/JHEP10(2023)022} {\bibfield  {journal} {\bibinfo  {journal} {JHEP}\
  }\textbf {\bibinfo {volume} {10}},\ \bibinfo {pages} {022} (\bibinfo {year}
  {2023})},\ \Eprint {http://arxiv.org/abs/2210.14865} {arXiv:2210.14865
  [hep-th]} \BibitemShut {NoStop}%
\bibitem [{\citenamefont {Bourjaily}\ \emph {et~al.}(2020)\citenamefont
  {Bourjaily}, \citenamefont {Herrmann}, \citenamefont {Langer},\ and\
  \citenamefont {Trnka}}]{Bourjaily:2020qca}%
  \BibitemOpen
  \bibfield  {author} {\bibinfo {author} {\bibfnamefont {Jacob~L.}\
  \bibnamefont {Bourjaily}}, \bibinfo {author} {\bibfnamefont {Enrico}\
  \bibnamefont {Herrmann}}, \bibinfo {author} {\bibfnamefont {Cameron}\
  \bibnamefont {Langer}}, \ and\ \bibinfo {author} {\bibfnamefont {Jaroslav}\
  \bibnamefont {Trnka}},\ }\bibfield  {title} {\enquote {\bibinfo {title}
  {{Building bases of loop integrands}},}\ }\href {\doibase
  10.1007/JHEP11(2020)116} {\bibfield  {journal} {\bibinfo  {journal} {JHEP}\
  }\textbf {\bibinfo {volume} {11}},\ \bibinfo {pages} {116} (\bibinfo {year}
  {2020})},\ \Eprint {http://arxiv.org/abs/2007.13905} {arXiv:2007.13905
  [hep-th]} \BibitemShut {NoStop}%
\bibitem [{\citenamefont {Bourjaily}\ \emph {et~al.}(2017)\citenamefont
  {Bourjaily}, \citenamefont {Herrmann},\ and\ \citenamefont
  {Trnka}}]{Bourjaily:2017wjl}%
  \BibitemOpen
  \bibfield  {author} {\bibinfo {author} {\bibfnamefont {Jacob~L.}\
  \bibnamefont {Bourjaily}}, \bibinfo {author} {\bibfnamefont {Enrico}\
  \bibnamefont {Herrmann}}, \ and\ \bibinfo {author} {\bibfnamefont {Jaroslav}\
  \bibnamefont {Trnka}},\ }\bibfield  {title} {\enquote {\bibinfo {title}
  {{Prescriptive Unitarity}},}\ }\href {\doibase 10.1007/JHEP06(2017)059}
  {\bibfield  {journal} {\bibinfo  {journal} {JHEP}\ }\textbf {\bibinfo
  {volume} {06}},\ \bibinfo {pages} {059} (\bibinfo {year} {2017})},\ \Eprint
  {http://arxiv.org/abs/1704.05460} {arXiv:1704.05460 [hep-th]} \BibitemShut
  {NoStop}%
\bibitem [{\citenamefont {Arkani-Hamed}\ \emph {et~al.}(2016)\citenamefont
  {Arkani-Hamed}, \citenamefont {Bourjaily}, \citenamefont {Cachazo},
  \citenamefont {Goncharov}, \citenamefont {Postnikov},\ and\ \citenamefont
  {Trnka}}]{Arkani-Hamed:2012zlh}%
  \BibitemOpen
  \bibfield  {author} {\bibinfo {author} {\bibfnamefont {Nima}\ \bibnamefont
  {Arkani-Hamed}}, \bibinfo {author} {\bibfnamefont {Jacob~L.}\ \bibnamefont
  {Bourjaily}}, \bibinfo {author} {\bibfnamefont {Freddy}\ \bibnamefont
  {Cachazo}}, \bibinfo {author} {\bibfnamefont {Alexander~B.}\ \bibnamefont
  {Goncharov}}, \bibinfo {author} {\bibfnamefont {Alexander}\ \bibnamefont
  {Postnikov}}, \ and\ \bibinfo {author} {\bibfnamefont {Jaroslav}\
  \bibnamefont {Trnka}},\ }\href {\doibase 10.1017/CBO9781316091548} {\emph
  {\bibinfo {title} {{Grassmannian Geometry of Scattering Amplitudes}}}}\
  (\bibinfo  {publisher} {Cambridge University Press},\ \bibinfo {year}
  {2016})\ \Eprint {http://arxiv.org/abs/1212.5605} {arXiv:1212.5605 [hep-th]}
  \BibitemShut {NoStop}%
\bibitem [{\citenamefont {Arkani-Hamed}\ and\ \citenamefont
  {Trnka}(2014)}]{Arkani-Hamed:2013jha}%
  \BibitemOpen
  \bibfield  {author} {\bibinfo {author} {\bibfnamefont {Nima}\ \bibnamefont
  {Arkani-Hamed}}\ and\ \bibinfo {author} {\bibfnamefont {Jaroslav}\
  \bibnamefont {Trnka}},\ }\bibfield  {title} {\enquote {\bibinfo {title} {{The
  Amplituhedron}},}\ }\href {\doibase 10.1007/JHEP10(2014)030} {\bibfield
  {journal} {\bibinfo  {journal} {JHEP}\ }\textbf {\bibinfo {volume} {10}},\
  \bibinfo {pages} {030} (\bibinfo {year} {2014})},\ \Eprint
  {http://arxiv.org/abs/1312.2007} {arXiv:1312.2007 [hep-th]} \BibitemShut
  {NoStop}%
\bibitem [{\citenamefont {Arkani-Hamed}\ \emph {et~al.}(2015)\citenamefont
  {Arkani-Hamed}, \citenamefont {Hodges},\ and\ \citenamefont
  {Trnka}}]{Arkani-Hamed:2014dca}%
  \BibitemOpen
  \bibfield  {author} {\bibinfo {author} {\bibfnamefont {Nima}\ \bibnamefont
  {Arkani-Hamed}}, \bibinfo {author} {\bibfnamefont {Andrew}\ \bibnamefont
  {Hodges}}, \ and\ \bibinfo {author} {\bibfnamefont {Jaroslav}\ \bibnamefont
  {Trnka}},\ }\bibfield  {title} {\enquote {\bibinfo {title} {{Positive
  Amplitudes In The Amplituhedron}},}\ }\href {\doibase
  10.1007/JHEP08(2015)030} {\bibfield  {journal} {\bibinfo  {journal} {JHEP}\
  }\textbf {\bibinfo {volume} {08}},\ \bibinfo {pages} {030} (\bibinfo {year}
  {2015})},\ \Eprint {http://arxiv.org/abs/1412.8478} {arXiv:1412.8478
  [hep-th]} \BibitemShut {NoStop}%
\bibitem [{\citenamefont {Drummond}\ \emph {et~al.}(2010)\citenamefont
  {Drummond}, \citenamefont {Henn}, \citenamefont {Korchemsky},\ and\
  \citenamefont {Sokatchev}}]{Drummond:2008vq}%
  \BibitemOpen
  \bibfield  {author} {\bibinfo {author} {\bibfnamefont {J.~M.}\ \bibnamefont
  {Drummond}}, \bibinfo {author} {\bibfnamefont {J.}~\bibnamefont {Henn}},
  \bibinfo {author} {\bibfnamefont {G.~P.}\ \bibnamefont {Korchemsky}}, \ and\
  \bibinfo {author} {\bibfnamefont {E.}~\bibnamefont {Sokatchev}},\ }\bibfield
  {title} {\enquote {\bibinfo {title} {{Dual superconformal symmetry of
  scattering amplitudes in N=4 super-Yang-Mills theory}},}\ }\href {\doibase
  10.1016/j.nuclphysb.2009.11.022} {\bibfield  {journal} {\bibinfo  {journal}
  {Nucl. Phys. B}\ }\textbf {\bibinfo {volume} {828}},\ \bibinfo {pages}
  {317--374} (\bibinfo {year} {2010})},\ \Eprint
  {http://arxiv.org/abs/0807.1095} {arXiv:0807.1095 [hep-th]} \BibitemShut
  {NoStop}%
\bibitem [{\citenamefont {Bourjaily}\ \emph {et~al.}(2016)\citenamefont
  {Bourjaily}, \citenamefont {Heslop},\ and\ \citenamefont
  {Tran}}]{Bourjaily:2016evz}%
  \BibitemOpen
  \bibfield  {author} {\bibinfo {author} {\bibfnamefont {Jacob~L.}\
  \bibnamefont {Bourjaily}}, \bibinfo {author} {\bibfnamefont {Paul}\
  \bibnamefont {Heslop}}, \ and\ \bibinfo {author} {\bibfnamefont {Vuong-Viet}\
  \bibnamefont {Tran}},\ }\bibfield  {title} {\enquote {\bibinfo {title}
  {{Amplitudes and Correlators to Ten Loops Using Simple, Graphical
  Bootstraps}},}\ }\href {\doibase 10.1007/JHEP11(2016)125} {\bibfield
  {journal} {\bibinfo  {journal} {JHEP}\ }\textbf {\bibinfo {volume} {11}},\
  \bibinfo {pages} {125} (\bibinfo {year} {2016})},\ \Eprint
  {http://arxiv.org/abs/1609.00007} {arXiv:1609.00007 [hep-th]} \BibitemShut
  {NoStop}%
\bibitem [{\citenamefont {Bern}\ \emph {et~al.}(2013)\citenamefont {Bern},
  \citenamefont {Carrasco}, \citenamefont {Dixon}, \citenamefont {Douglas},
  \citenamefont {von Hippel},\ and\ \citenamefont {Johansson}}]{Bern:2012di}%
  \BibitemOpen
  \bibfield  {author} {\bibinfo {author} {\bibfnamefont {Zvi}\ \bibnamefont
  {Bern}}, \bibinfo {author} {\bibfnamefont {John~Joseph}\ \bibnamefont
  {Carrasco}}, \bibinfo {author} {\bibfnamefont {Lance~J.}\ \bibnamefont
  {Dixon}}, \bibinfo {author} {\bibfnamefont {Michael~R.}\ \bibnamefont
  {Douglas}}, \bibinfo {author} {\bibfnamefont {Matt}\ \bibnamefont {von
  Hippel}}, \ and\ \bibinfo {author} {\bibfnamefont {Henrik}\ \bibnamefont
  {Johansson}},\ }\bibfield  {title} {\enquote {\bibinfo {title} {{D=5
  maximally supersymmetric Yang-Mills theory diverges at six loops}},}\ }\href
  {\doibase 10.1103/PhysRevD.87.025018} {\bibfield  {journal} {\bibinfo
  {journal} {Phys. Rev. D}\ }\textbf {\bibinfo {volume} {87}},\ \bibinfo
  {pages} {025018} (\bibinfo {year} {2013})},\ \Eprint
  {http://arxiv.org/abs/1210.7709} {arXiv:1210.7709 [hep-th]} \BibitemShut
  {NoStop}%
\bibitem [{\citenamefont {Dixon}\ \emph {et~al.}(2021)\citenamefont {Dixon},
  \citenamefont {Liu},\ and\ \citenamefont {Miczajka}}]{Dixon:2021nzr}%
  \BibitemOpen
  \bibfield  {author} {\bibinfo {author} {\bibfnamefont {Lance~J.}\
  \bibnamefont {Dixon}}, \bibinfo {author} {\bibfnamefont {Yu-Ting}\
  \bibnamefont {Liu}}, \ and\ \bibinfo {author} {\bibfnamefont {Julian}\
  \bibnamefont {Miczajka}},\ }\bibfield  {title} {\enquote {\bibinfo {title}
  {{Heptagon functions and seven-gluon amplitudes in multi-Regge
  kinematics}},}\ }\href {\doibase 10.1007/JHEP12(2021)218} {\bibfield
  {journal} {\bibinfo  {journal} {JHEP}\ }\textbf {\bibinfo {volume} {12}},\
  \bibinfo {pages} {218} (\bibinfo {year} {2021})},\ \Eprint
  {http://arxiv.org/abs/2110.11388} {arXiv:2110.11388 [hep-th]} \BibitemShut
  {NoStop}%
\bibitem [{\citenamefont {Caron-Huot}\ \emph {et~al.}(2019)\citenamefont
  {Caron-Huot}, \citenamefont {Dixon}, \citenamefont {Dulat}, \citenamefont
  {von Hippel}, \citenamefont {McLeod},\ and\ \citenamefont
  {Papathanasiou}}]{Caron-Huot:2019vjl}%
  \BibitemOpen
  \bibfield  {author} {\bibinfo {author} {\bibfnamefont {Simon}\ \bibnamefont
  {Caron-Huot}}, \bibinfo {author} {\bibfnamefont {Lance~J.}\ \bibnamefont
  {Dixon}}, \bibinfo {author} {\bibfnamefont {Falko}\ \bibnamefont {Dulat}},
  \bibinfo {author} {\bibfnamefont {Matt}\ \bibnamefont {von Hippel}}, \bibinfo
  {author} {\bibfnamefont {Andrew~J.}\ \bibnamefont {McLeod}}, \ and\ \bibinfo
  {author} {\bibfnamefont {Georgios}\ \bibnamefont {Papathanasiou}},\
  }\bibfield  {title} {\enquote {\bibinfo {title} {{Six-Gluon amplitudes in
  planar $ \mathcal{N} $ = 4 super-Yang-Mills theory at six and seven
  loops}},}\ }\href {\doibase 10.1007/JHEP08(2019)016} {\bibfield  {journal}
  {\bibinfo  {journal} {JHEP}\ }\textbf {\bibinfo {volume} {08}},\ \bibinfo
  {pages} {016} (\bibinfo {year} {2019})},\ \Eprint
  {http://arxiv.org/abs/1903.10890} {arXiv:1903.10890 [hep-th]} \BibitemShut
  {NoStop}%
\bibitem [{\citenamefont {Bern}\ \emph {et~al.}(2008)\citenamefont {Bern},
  \citenamefont {Carrasco},\ and\ \citenamefont {Johansson}}]{Bern:2008qj}%
  \BibitemOpen
  \bibfield  {author} {\bibinfo {author} {\bibfnamefont {Z.}~\bibnamefont
  {Bern}}, \bibinfo {author} {\bibfnamefont {J.~J.~M.}\ \bibnamefont
  {Carrasco}}, \ and\ \bibinfo {author} {\bibfnamefont {Henrik}\ \bibnamefont
  {Johansson}},\ }\bibfield  {title} {\enquote {\bibinfo {title} {{New
  Relations for Gauge-Theory Amplitudes}},}\ }\href {\doibase
  10.1103/PhysRevD.78.085011} {\bibfield  {journal} {\bibinfo  {journal} {Phys.
  Rev. D}\ }\textbf {\bibinfo {volume} {78}},\ \bibinfo {pages} {085011}
  (\bibinfo {year} {2008})},\ \Eprint {http://arxiv.org/abs/0805.3993}
  {arXiv:0805.3993 [hep-ph]} \BibitemShut {NoStop}%
\bibitem [{\citenamefont {Kawai}\ \emph {et~al.}(1986)\citenamefont {Kawai},
  \citenamefont {Lewellen},\ and\ \citenamefont {Tye}}]{Kawai:1985xq}%
  \BibitemOpen
  \bibfield  {author} {\bibinfo {author} {\bibfnamefont {H.}~\bibnamefont
  {Kawai}}, \bibinfo {author} {\bibfnamefont {D.~C.}\ \bibnamefont {Lewellen}},
  \ and\ \bibinfo {author} {\bibfnamefont {S.~H.~H.}\ \bibnamefont {Tye}},\
  }\bibfield  {title} {\enquote {\bibinfo {title} {{A Relation Between Tree
  Amplitudes of Closed and Open Strings}},}\ }\href {\doibase
  10.1016/0550-3213(86)90362-7} {\bibfield  {journal} {\bibinfo  {journal}
  {Nucl. Phys. B}\ }\textbf {\bibinfo {volume} {269}},\ \bibinfo {pages}
  {1--23} (\bibinfo {year} {1986})}\BibitemShut {NoStop}%
\bibitem [{\citenamefont {Bern}\ \emph {et~al.}(2010)\citenamefont {Bern},
  \citenamefont {Carrasco},\ and\ \citenamefont {Johansson}}]{Bern:2010ue}%
  \BibitemOpen
  \bibfield  {author} {\bibinfo {author} {\bibfnamefont {Zvi}\ \bibnamefont
  {Bern}}, \bibinfo {author} {\bibfnamefont {John Joseph~M.}\ \bibnamefont
  {Carrasco}}, \ and\ \bibinfo {author} {\bibfnamefont {Henrik}\ \bibnamefont
  {Johansson}},\ }\bibfield  {title} {\enquote {\bibinfo {title} {{Perturbative
  Quantum Gravity as a Double Copy of Gauge Theory}},}\ }\href {\doibase
  10.1103/PhysRevLett.105.061602} {\bibfield  {journal} {\bibinfo  {journal}
  {Phys. Rev. Lett.}\ }\textbf {\bibinfo {volume} {105}},\ \bibinfo {pages}
  {061602} (\bibinfo {year} {2010})},\ \Eprint {http://arxiv.org/abs/1004.0476}
  {arXiv:1004.0476 [hep-th]} \BibitemShut {NoStop}%
\bibitem [{\citenamefont {Bern}\ \emph {et~al.}(2017)\citenamefont {Bern},
  \citenamefont {Carrasco}, \citenamefont {Chen}, \citenamefont {Johansson},\
  and\ \citenamefont {Roiban}}]{Bern:2017yxu}%
  \BibitemOpen
  \bibfield  {author} {\bibinfo {author} {\bibfnamefont {Zvi}\ \bibnamefont
  {Bern}}, \bibinfo {author} {\bibfnamefont {John~Joseph}\ \bibnamefont
  {Carrasco}}, \bibinfo {author} {\bibfnamefont {Wei-Ming}\ \bibnamefont
  {Chen}}, \bibinfo {author} {\bibfnamefont {Henrik}\ \bibnamefont
  {Johansson}}, \ and\ \bibinfo {author} {\bibfnamefont {Radu}\ \bibnamefont
  {Roiban}},\ }\bibfield  {title} {\enquote {\bibinfo {title} {{Gravity
  Amplitudes as Generalized Double Copies of Gauge-Theory Amplitudes}},}\
  }\href {\doibase 10.1103/PhysRevLett.118.181602} {\bibfield  {journal}
  {\bibinfo  {journal} {Phys. Rev. Lett.}\ }\textbf {\bibinfo {volume} {118}},\
  \bibinfo {pages} {181602} (\bibinfo {year} {2017})},\ \Eprint
  {http://arxiv.org/abs/1701.02519} {arXiv:1701.02519 [hep-th]} \BibitemShut
  {NoStop}%
\bibitem [{\citenamefont {Cachazo}\ \emph {et~al.}(2015)\citenamefont
  {Cachazo}, \citenamefont {He},\ and\ \citenamefont {Yuan}}]{Cachazo:2014xea}%
  \BibitemOpen
  \bibfield  {author} {\bibinfo {author} {\bibfnamefont {Freddy}\ \bibnamefont
  {Cachazo}}, \bibinfo {author} {\bibfnamefont {Song}\ \bibnamefont {He}}, \
  and\ \bibinfo {author} {\bibfnamefont {Ellis~Ye}\ \bibnamefont {Yuan}},\
  }\bibfield  {title} {\enquote {\bibinfo {title} {{Scattering Equations and
  Matrices: From Einstein To Yang-Mills, DBI and NLSM}},}\ }\href {\doibase
  10.1007/JHEP07(2015)149} {\bibfield  {journal} {\bibinfo  {journal} {JHEP}\
  }\textbf {\bibinfo {volume} {07}},\ \bibinfo {pages} {149} (\bibinfo {year}
  {2015})},\ \Eprint {http://arxiv.org/abs/1412.3479} {arXiv:1412.3479
  [hep-th]} \BibitemShut {NoStop}%
\bibitem [{\citenamefont {Bern}\ \emph {et~al.}(2019)\citenamefont {Bern},
  \citenamefont {Carrasco}, \citenamefont {Chiodaroli}, \citenamefont
  {Johansson},\ and\ \citenamefont {Roiban}}]{Bern:2019prr}%
  \BibitemOpen
  \bibfield  {author} {\bibinfo {author} {\bibfnamefont {Zvi}\ \bibnamefont
  {Bern}}, \bibinfo {author} {\bibfnamefont {John~Joseph}\ \bibnamefont
  {Carrasco}}, \bibinfo {author} {\bibfnamefont {Marco}\ \bibnamefont
  {Chiodaroli}}, \bibinfo {author} {\bibfnamefont {Henrik}\ \bibnamefont
  {Johansson}}, \ and\ \bibinfo {author} {\bibfnamefont {Radu}\ \bibnamefont
  {Roiban}},\ }\bibfield  {title} {\enquote {\bibinfo {title} {{The Duality
  Between Color and Kinematics and its Applications}},}\ }\href@noop {} {\
  (\bibinfo {year} {2019})},\ \Eprint {http://arxiv.org/abs/1909.01358}
  {arXiv:1909.01358 [hep-th]} \BibitemShut {NoStop}%
\bibitem [{\citenamefont {Adamo}\ \emph {et~al.}(2022)\citenamefont {Adamo},
  \citenamefont {Carrasco}, \citenamefont {Carrillo-Gonz\'alez}, \citenamefont
  {Chiodaroli}, \citenamefont {Elvang}, \citenamefont {Johansson},
  \citenamefont {O'Connell}, \citenamefont {Roiban},\ and\ \citenamefont
  {Schlotterer}}]{Adamo:2022dcm}%
  \BibitemOpen
  \bibfield  {author} {\bibinfo {author} {\bibfnamefont {Tim}\ \bibnamefont
  {Adamo}}, \bibinfo {author} {\bibfnamefont {John Joseph~M.}\ \bibnamefont
  {Carrasco}}, \bibinfo {author} {\bibfnamefont {Mariana}\ \bibnamefont
  {Carrillo-Gonz\'alez}}, \bibinfo {author} {\bibfnamefont {Marco}\
  \bibnamefont {Chiodaroli}}, \bibinfo {author} {\bibfnamefont {Henriette}\
  \bibnamefont {Elvang}}, \bibinfo {author} {\bibfnamefont {Henrik}\
  \bibnamefont {Johansson}}, \bibinfo {author} {\bibfnamefont {Donal}\
  \bibnamefont {O'Connell}}, \bibinfo {author} {\bibfnamefont {Radu}\
  \bibnamefont {Roiban}}, \ and\ \bibinfo {author} {\bibfnamefont {Oliver}\
  \bibnamefont {Schlotterer}},\ }\bibfield  {title} {\enquote {\bibinfo {title}
  {{Snowmass White Paper: the Double Copy and its Applications}},}\ }in\
  \href@noop {} {\emph {\bibinfo {booktitle} {{Snowmass 2021}}}}\ (\bibinfo
  {year} {2022})\ \Eprint {http://arxiv.org/abs/2204.06547} {arXiv:2204.06547
  [hep-th]} \BibitemShut {NoStop}%
\bibitem [{\citenamefont {Monteiro}\ \emph {et~al.}(2014)\citenamefont
  {Monteiro}, \citenamefont {O'Connell},\ and\ \citenamefont
  {White}}]{Monteiro:2014cda}%
  \BibitemOpen
  \bibfield  {author} {\bibinfo {author} {\bibfnamefont {Ricardo}\ \bibnamefont
  {Monteiro}}, \bibinfo {author} {\bibfnamefont {Donal}\ \bibnamefont
  {O'Connell}}, \ and\ \bibinfo {author} {\bibfnamefont {Chris~D.}\
  \bibnamefont {White}},\ }\bibfield  {title} {\enquote {\bibinfo {title}
  {{Black holes and the double copy}},}\ }\href {\doibase
  10.1007/JHEP12(2014)056} {\bibfield  {journal} {\bibinfo  {journal} {JHEP}\
  }\textbf {\bibinfo {volume} {12}},\ \bibinfo {pages} {056} (\bibinfo {year}
  {2014})},\ \Eprint {http://arxiv.org/abs/1410.0239} {arXiv:1410.0239
  [hep-th]} \BibitemShut {NoStop}%
\bibitem [{\citenamefont {Kosower}\ \emph {et~al.}(2019)\citenamefont
  {Kosower}, \citenamefont {Maybee},\ and\ \citenamefont
  {O'Connell}}]{Kosower:2018adc}%
  \BibitemOpen
  \bibfield  {author} {\bibinfo {author} {\bibfnamefont {David~A.}\
  \bibnamefont {Kosower}}, \bibinfo {author} {\bibfnamefont {Ben}\ \bibnamefont
  {Maybee}}, \ and\ \bibinfo {author} {\bibfnamefont {Donal}\ \bibnamefont
  {O'Connell}},\ }\bibfield  {title} {\enquote {\bibinfo {title} {{Amplitudes,
  Observables, and Classical Scattering}},}\ }\href {\doibase
  10.1007/JHEP02(2019)137} {\bibfield  {journal} {\bibinfo  {journal} {JHEP}\
  }\textbf {\bibinfo {volume} {02}},\ \bibinfo {pages} {137} (\bibinfo {year}
  {2019})},\ \Eprint {http://arxiv.org/abs/1811.10950} {arXiv:1811.10950
  [hep-th]} \BibitemShut {NoStop}%
\bibitem [{\citenamefont {Buonanno}\ \emph {et~al.}(2022)\citenamefont
  {Buonanno}, \citenamefont {Khalil}, \citenamefont {O'Connell}, \citenamefont
  {Roiban}, \citenamefont {Solon},\ and\ \citenamefont
  {Zeng}}]{Buonanno:2022pgc}%
  \BibitemOpen
  \bibfield  {author} {\bibinfo {author} {\bibfnamefont {Alessandra}\
  \bibnamefont {Buonanno}}, \bibinfo {author} {\bibfnamefont {Mohammed}\
  \bibnamefont {Khalil}}, \bibinfo {author} {\bibfnamefont {Donal}\
  \bibnamefont {O'Connell}}, \bibinfo {author} {\bibfnamefont {Radu}\
  \bibnamefont {Roiban}}, \bibinfo {author} {\bibfnamefont {Mikhail~P.}\
  \bibnamefont {Solon}}, \ and\ \bibinfo {author} {\bibfnamefont {Mao}\
  \bibnamefont {Zeng}},\ }\bibfield  {title} {\enquote {\bibinfo {title}
  {{Snowmass White Paper: Gravitational Waves and Scattering Amplitudes}},}\
  }in\ \href@noop {} {\emph {\bibinfo {booktitle} {{Snowmass 2021}}}}\
  (\bibinfo {year} {2022})\ \Eprint {http://arxiv.org/abs/2204.05194}
  {arXiv:2204.05194 [hep-th]} \BibitemShut {NoStop}%
\bibitem [{\citenamefont {Bern}\ \emph
  {et~al.}(2016{\natexlab{a}})\citenamefont {Bern}, \citenamefont {Davies},\
  and\ \citenamefont {Nohle}}]{Bern:2015ooa}%
  \BibitemOpen
  \bibfield  {author} {\bibinfo {author} {\bibfnamefont {Zvi}\ \bibnamefont
  {Bern}}, \bibinfo {author} {\bibfnamefont {Scott}\ \bibnamefont {Davies}}, \
  and\ \bibinfo {author} {\bibfnamefont {Josh}\ \bibnamefont {Nohle}},\
  }\bibfield  {title} {\enquote {\bibinfo {title} {{Double-Copy Constructions
  and Unitarity Cuts}},}\ }\href {\doibase 10.1103/PhysRevD.93.105015}
  {\bibfield  {journal} {\bibinfo  {journal} {Phys. Rev. D}\ }\textbf {\bibinfo
  {volume} {93}},\ \bibinfo {pages} {105015} (\bibinfo {year}
  {2016}{\natexlab{a}})},\ \Eprint {http://arxiv.org/abs/1510.03448}
  {arXiv:1510.03448 [hep-th]} \BibitemShut {NoStop}%
\bibitem [{\citenamefont {Mogull}\ and\ \citenamefont
  {O'Connell}(2015)}]{Mogull:2015adi}%
  \BibitemOpen
  \bibfield  {author} {\bibinfo {author} {\bibfnamefont {Gustav}\ \bibnamefont
  {Mogull}}\ and\ \bibinfo {author} {\bibfnamefont {Donal}\ \bibnamefont
  {O'Connell}},\ }\bibfield  {title} {\enquote {\bibinfo {title} {{Overcoming
  Obstacles to Colour-Kinematics Duality at Two Loops}},}\ }\href {\doibase
  10.1007/JHEP12(2015)135} {\bibfield  {journal} {\bibinfo  {journal} {JHEP}\
  }\textbf {\bibinfo {volume} {12}},\ \bibinfo {pages} {135} (\bibinfo {year}
  {2015})},\ \Eprint {http://arxiv.org/abs/1511.06652} {arXiv:1511.06652
  [hep-th]} \BibitemShut {NoStop}%
\bibitem [{\citenamefont {Edison}\ \emph {et~al.}(2024)\citenamefont {Edison},
  \citenamefont {Mangan},\ and\ \citenamefont {Pavao}}]{Edison:2023ulf}%
  \BibitemOpen
  \bibfield  {author} {\bibinfo {author} {\bibfnamefont {Alex}\ \bibnamefont
  {Edison}}, \bibinfo {author} {\bibfnamefont {James}\ \bibnamefont {Mangan}},
  \ and\ \bibinfo {author} {\bibfnamefont {Nicolas~H.}\ \bibnamefont {Pavao}},\
  }\bibfield  {title} {\enquote {\bibinfo {title} {{Revealing the landscape of
  globally color-dual multi-loop integrands}},}\ }\href {\doibase
  10.1007/JHEP03(2024)163} {\bibfield  {journal} {\bibinfo  {journal} {JHEP}\
  }\textbf {\bibinfo {volume} {03}},\ \bibinfo {pages} {163} (\bibinfo {year}
  {2024})},\ \Eprint {http://arxiv.org/abs/2309.16558} {arXiv:2309.16558
  [hep-th]} \BibitemShut {NoStop}%
\bibitem [{\citenamefont {Li}\ and\ \citenamefont {Yang}(2024)}]{Li:2023akg}%
  \BibitemOpen
  \bibfield  {author} {\bibinfo {author} {\bibfnamefont {Zeyu}\ \bibnamefont
  {Li}}\ and\ \bibinfo {author} {\bibfnamefont {Gang}\ \bibnamefont {Yang}},\
  }\bibfield  {title} {\enquote {\bibinfo {title} {{Color-kinematics duality
  with minimal deformation: two-loop four-gluon amplitudes in pure Yang-Mills
  revisited}},}\ }\href {\doibase 10.1007/JHEP02(2024)199} {\bibfield
  {journal} {\bibinfo  {journal} {JHEP}\ }\textbf {\bibinfo {volume} {02}},\
  \bibinfo {pages} {199} (\bibinfo {year} {2024})},\ \Eprint
  {http://arxiv.org/abs/2312.04319} {arXiv:2312.04319 [hep-th]} \BibitemShut
  {NoStop}%
\bibitem [{\citenamefont {Cachazo}\ \emph {et~al.}(2014)\citenamefont
  {Cachazo}, \citenamefont {He},\ and\ \citenamefont {Yuan}}]{Cachazo:2013hca}%
  \BibitemOpen
  \bibfield  {author} {\bibinfo {author} {\bibfnamefont {Freddy}\ \bibnamefont
  {Cachazo}}, \bibinfo {author} {\bibfnamefont {Song}\ \bibnamefont {He}}, \
  and\ \bibinfo {author} {\bibfnamefont {Ellis~Ye}\ \bibnamefont {Yuan}},\
  }\bibfield  {title} {\enquote {\bibinfo {title} {{Scattering of Massless
  Particles in Arbitrary Dimensions}},}\ }\href {\doibase
  10.1103/PhysRevLett.113.171601} {\bibfield  {journal} {\bibinfo  {journal}
  {Phys. Rev. Lett.}\ }\textbf {\bibinfo {volume} {113}},\ \bibinfo {pages}
  {171601} (\bibinfo {year} {2014})},\ \Eprint {http://arxiv.org/abs/1307.2199}
  {arXiv:1307.2199 [hep-th]} \BibitemShut {NoStop}%
\bibitem [{\citenamefont {Carrasco}\ \emph {et~al.}(2017)\citenamefont
  {Carrasco}, \citenamefont {Mafra},\ and\ \citenamefont
  {Schlotterer}}]{Carrasco:2016ldy}%
  \BibitemOpen
  \bibfield  {author} {\bibinfo {author} {\bibfnamefont {John Joseph~M.}\
  \bibnamefont {Carrasco}}, \bibinfo {author} {\bibfnamefont {Carlos~R.}\
  \bibnamefont {Mafra}}, \ and\ \bibinfo {author} {\bibfnamefont {Oliver}\
  \bibnamefont {Schlotterer}},\ }\bibfield  {title} {\enquote {\bibinfo {title}
  {{Abelian Z-theory: NLSM amplitudes and $\alpha$'-corrections from the open
  string}},}\ }\href {\doibase 10.1007/JHEP06(2017)093} {\bibfield  {journal}
  {\bibinfo  {journal} {JHEP}\ }\textbf {\bibinfo {volume} {06}},\ \bibinfo
  {pages} {093} (\bibinfo {year} {2017})},\ \Eprint
  {http://arxiv.org/abs/1608.02569} {arXiv:1608.02569 [hep-th]} \BibitemShut
  {NoStop}%
\bibitem [{\citenamefont {Kleiss}\ and\ \citenamefont
  {Kuijf}(1989)}]{Kleiss:1988ne}%
  \BibitemOpen
  \bibfield  {author} {\bibinfo {author} {\bibfnamefont {Ronald}\ \bibnamefont
  {Kleiss}}\ and\ \bibinfo {author} {\bibfnamefont {Hans}\ \bibnamefont
  {Kuijf}},\ }\bibfield  {title} {\enquote {\bibinfo {title} {{Multi - Gluon
  Cross-sections and Five Jet Production at Hadron Colliders}},}\ }\href
  {\doibase 10.1016/0550-3213(89)90574-9} {\bibfield  {journal} {\bibinfo
  {journal} {Nucl. Phys. B}\ }\textbf {\bibinfo {volume} {312}},\ \bibinfo
  {pages} {616--644} (\bibinfo {year} {1989})}\BibitemShut {NoStop}%
\bibitem [{\citenamefont {Bern}\ \emph
  {et~al.}(2016{\natexlab{b}})\citenamefont {Bern}, \citenamefont {Herrmann},
  \citenamefont {Litsey}, \citenamefont {Stankowicz},\ and\ \citenamefont
  {Trnka}}]{Bern:2015ple}%
  \BibitemOpen
  \bibfield  {author} {\bibinfo {author} {\bibfnamefont {Zvi}\ \bibnamefont
  {Bern}}, \bibinfo {author} {\bibfnamefont {Enrico}\ \bibnamefont {Herrmann}},
  \bibinfo {author} {\bibfnamefont {Sean}\ \bibnamefont {Litsey}}, \bibinfo
  {author} {\bibfnamefont {James}\ \bibnamefont {Stankowicz}}, \ and\ \bibinfo
  {author} {\bibfnamefont {Jaroslav}\ \bibnamefont {Trnka}},\ }\bibfield
  {title} {\enquote {\bibinfo {title} {{Evidence for a Nonplanar
  Amplituhedron}},}\ }\href {\doibase 10.1007/JHEP06(2016)098} {\bibfield
  {journal} {\bibinfo  {journal} {JHEP}\ }\textbf {\bibinfo {volume} {06}},\
  \bibinfo {pages} {098} (\bibinfo {year} {2016}{\natexlab{b}})},\ \Eprint
  {http://arxiv.org/abs/1512.08591} {arXiv:1512.08591 [hep-th]} \BibitemShut
  {NoStop}%
\bibitem [{\citenamefont {Edison}\ \emph {et~al.}(2021)\citenamefont {Edison},
  \citenamefont {Herrmann}, \citenamefont {Parra-Martinez},\ and\ \citenamefont
  {Trnka}}]{Edison:2019ovj}%
  \BibitemOpen
  \bibfield  {author} {\bibinfo {author} {\bibfnamefont {Alex}\ \bibnamefont
  {Edison}}, \bibinfo {author} {\bibfnamefont {Enrico}\ \bibnamefont
  {Herrmann}}, \bibinfo {author} {\bibfnamefont {Julio}\ \bibnamefont
  {Parra-Martinez}}, \ and\ \bibinfo {author} {\bibfnamefont {Jaroslav}\
  \bibnamefont {Trnka}},\ }\bibfield  {title} {\enquote {\bibinfo {title}
  {{Gravity loop integrands from the ultraviolet}},}\ }\href {\doibase
  10.21468/SciPostPhys.10.1.016} {\bibfield  {journal} {\bibinfo  {journal}
  {SciPost Phys.}\ }\textbf {\bibinfo {volume} {10}},\ \bibinfo {pages} {016}
  (\bibinfo {year} {2021})},\ \Eprint {http://arxiv.org/abs/1909.02003}
  {arXiv:1909.02003 [hep-th]} \BibitemShut {NoStop}%
\bibitem [{\citenamefont {Bern}\ \emph {et~al.}(2024)\citenamefont {Bern},
  \citenamefont {Herrmann}, \citenamefont {Roiban}, \citenamefont {Ruf},\ and\
  \citenamefont {Zeng}}]{Bern:2024vqs}%
  \BibitemOpen
  \bibfield  {author} {\bibinfo {author} {\bibfnamefont {Zvi}\ \bibnamefont
  {Bern}}, \bibinfo {author} {\bibfnamefont {Enrico}\ \bibnamefont {Herrmann}},
  \bibinfo {author} {\bibfnamefont {Radu}\ \bibnamefont {Roiban}}, \bibinfo
  {author} {\bibfnamefont {Michael~S.}\ \bibnamefont {Ruf}}, \ and\ \bibinfo
  {author} {\bibfnamefont {Mao}\ \bibnamefont {Zeng}},\ }\bibfield  {title}
  {\enquote {\bibinfo {title} {{Global Bases for Nonplanar Loop Integrands,
  Generalized Unitarity, and the Double Copy to All Loop Orders}},}\
  }\href@noop {} {\  (\bibinfo {year} {2024})},\ \Eprint
  {http://arxiv.org/abs/2408.06686} {arXiv:2408.06686 [hep-th]} \BibitemShut
  {NoStop}%
\bibitem [{\citenamefont {{Gerosa}}\ and\ \citenamefont
  {{Vallisneri}}(2017)}]{2017JOSS....2..222G}%
  \BibitemOpen
  \bibfield  {author} {\bibinfo {author} {\bibfnamefont {Davide}\ \bibnamefont
  {{Gerosa}}}\ and\ \bibinfo {author} {\bibfnamefont {Michele}\ \bibnamefont
  {{Vallisneri}}},\ }\bibfield  {title} {\enquote {\bibinfo {title} {{filltex:
  Automatic queries to ADS and INSPIRE databases to fill LaTex
  bibliography}},}\ }\href {\doibase 10.21105/joss.00222} {\bibfield  {journal}
  {\bibinfo  {journal} {The Journal of Open Source Software}\ }\textbf
  {\bibinfo {volume} {2}},\ \bibinfo {pages} {222} (\bibinfo {year}
  {2017})}\BibitemShut {NoStop}%
\end{thebibliography}%
